\documentclass[
draft=false, 
toc=bibliography 
]{scrreprt}
\usepackage{url}
\usepackage{hyperref}
\usepackage{booktabs} 
\usepackage{graphicx}
\usepackage{verbatim}
\usepackage{amsmath}
\usepackage{color}
\usepackage{makeidx}

\def\lsim{\mathrel{\rlap{\lower4pt\hbox{\hskip1pt$\sim$}}
    \raise1pt\hbox{$<$}}}                
\def\gsim{\mathrel{\rlap{\lower4pt\hbox{\hskip1pt$\sim$}}
    \raise1pt\hbox{$>$}}}                

\makeindex
\begin{document}
\subject{Reference manual\\version 1.00}
\title{CompOSE}
\subtitle{
	Comp{\rm Star} O{\rm nline} S{\rm upernov\ae} E{\rm quations of State}
	\\[5ex]
	{\it harmonising the concert of nuclear physics and astrophysics}
	\\[5ex]
	{\rm compose.obspm.fr
        }
	\\[10ex]
        \mbox{}
}
\author{CompOSE Core Team}
\date{\today}
\maketitle
\tableofcontents

\part{Introduction}
\chapter{What CompOSE can do and what not}
The online service CompOSE provides information and
data tables for different equations of 
states (EoS) ready for further use
in astrophysical applications,
nuclear physics and beyond.
This service has three major purposes:
\begin{itemize}
\item CompOSE is a repository of EoS tables in a common format
for direct usage with
information on a large number of
thermodynamic properties, on the chemical composition 
of dense matter and, if available, on
microphysical quantities of the constituents.
\item CompOSE allows to interpolate the provided tables  
using different schemes to obtain the relevant quantities, selected by
the  user, for grids that are tailored to specific applications.
\item CompOSE can provide information on
additional thermodynamic quantities, 
which are not stored in the original data tables, and on further
quantities,
which charactize an EoS such as nuclear matter parameters and compact
star properties.
\end{itemize} 
The format of the files as well as the calculational mesh is mainly determined 
according to the needs of scientific groups performing extensive numerical 
simulations of astrophysical objects.

We cannot offer an online service for all features of CompOSE,
e.g.\ to run all codes online.
This is mostly due to limitations in storage and computation times
but also gives better control on 
avoiding unphysical input parameters.
However, we offer several computational tools that allow the user
to extract the data from the tables that are relevant for her/him.
We try to provide the tables in a large parameter space to cover most
applications.

If you make use of the tables provided, you will be guided on the
CompOSE web pages to the scientific 
publications where the particular EoS models have been described in detail.
Please cite them when using the tables for scientific purposes.

CompOSE is designed in a modular way, thus allowing to extend the service over 
time.
More and more models and parameter sets will be provided in time.
It is foreseen that additional features will be added in the future, too.

\chapter{How to read this document}
\label{ch:document}
While reading this document please always keep in mind: this document was 
written by physicists for physicists.
It is divided into three major parts.

The first one is relevant both to providers and users of equations of state
since it serves as a basis for the discussions in the following parts
of the manual and the web site. 
Both, contributors and users, should first of all have a look at the 
introductory chapter \ref{ch:defnot}
where we will discuss general conventions and the notation used throughout
CompOSE. In addition, we give definitions and details on
the system of units that is used within CompOSE and 
on physical constants that should be used 
in order to standardise the generation of new equations of state.

The second part concerns those persons
who wish to contribute to an extension of the CompOSE data base
by the active development of an EoS. CompOSE allows them to make their
favourite EoS available for a broad range of astrophysical and nuclear
physics applications.
In part \ref{part:contrib} detailed instructions, minimal
requirements and recommendations are 
specified for the preparation of EoS tables
that can be incorporated in future versions of the CompOSE data base. 
If you plan to contribute your EoS, 
you should contact the CompOSE core team, 
see appendix \ref{app:team}.
A summary of possible future extensions of the CompOSE data base
are summarized in chapter \ref{ch:extensions}.

The third part concerns the users of EoS data, that are provided by
CompOSE,
who want to test various equations of state in their
simulations of core-collapse supernovae, neutron-star mergers and
other scenarios. In general, they can
safely skip the second part and go directly to the third one. 
This part gives a brief introduction on nuclear
matter properties relevant for the construction of an EoS as well as a
classification of different types of EoS models.
The models can be distinguished either by using
different techniques to treat the many-body system 
of strongly interacting particles
or by assuming a different particle composition. The main aim
is to give the relevant information for the
interpretation of the data sheets, provided with each
available EoS table on the web site.

On the latter
you will find remarks on 
the range of applicability of the various models,
i.e.\ the range of parameters where the code is tested and/or the 
made approximations are still valid.
Characteristic parameters of each model will be specified.
For more detailed information about the physics behind each model, we refer to
the original references. 
In addition, the third part of the manual
explains how to proceed in order to download an EoS table and
the computational tools. 
The latter allow in particular the generation of tables with a
mesh different from the original one via interpolation, in order to adapt
the table to the need of the user. In addition, several thermodynamic
quantities can be calculated which are not contained in the original tables.
The use of the online service and web interface is described in
chapter \ref{ch:online-service}.


\chapter{Definitions and notation}
\label{ch:defnot}

The equations of state in the CompOSE data base are provided under
some common assumptions on the physical conditions of the matter
that are specified in this chapter. In order to fix the notation in 
the present document, the definition of all relevant quantities is
given.

\section{Units and conventions}

We use natural units\index{units} 
$\hbar=c=k_{B}=1$\index{$\hbar$}\index{$c$}\index{$k_{B}$}
throughout this document as
is customary in nuclear physics.
Energies\index{energy} and temperatures\index{temperature} 
are measured in MeV, lengths\index{length} in fm.
Units are given in parentheses $[\dots]$ for all quantities
when they are defined within this chapter.
For conversion among your favourite units the use 
of the  CODATA\index{CODATA} values \cite{CODATA}
(\url{http://physics.nist.gov/cuu/Constants/index.html}, \url{www.codata.org})
is recommended, see 
Tab.~\ref{tab:codata}\footnote{The compilation 
on physical constants by the Particle Data Group,
\url{pdg.lbl.gov}, contains the CODATA values together with some additional
constants concerning the interaction of elementary particles.}.
They should also be used in the preparation
of new EoS tables to be incorporated in the CompOSE data base.

For experimental 
binding energies\index{binding energy} of nuclei, the values of the
2012 Atomic mass evaluation\index{Atomic Mass Evaluation} \cite{Ame2012}
(AME2012, \url{http://amdc.impcas.ac.cn})
or corresponding up\-dates are recommended.
Ground state spin\index{spin!ground state} assignments should be taken from
the Nubase2012\index{Nubase} \cite{Nubase2012}
evaluation of nuclear properties
(\url{http://amdc.impcas.ac.cn})
or corresponding updates.

\begin{table}[htdp]
\begin{center}
\caption{\label{tab:codata}%
Recommended values for physical constants from the 2010
CODATA evaluation. The
elementary charge is denoted by the symbol $e$.}
\begin{tabular}{llll}
\toprule
Quantity & Symbol & Value & Unit \\
\midrule
speed of light\index{speed of light} 
in vacuum & $c\index{$c$}$ & $299792458$ & m~s$^{-1}$
\\
Planck's constant\index{constant!Planck's} 
&$\hbar\index{$\hbar$}$ & $1.054571726 \times 10^{-34}$ & J~s\\
                 &        & $6.58211928 \times 10^{-22}$ & MeV~s \\
                 &        & $197.3269718$ & MeV~fm~$c^{-1}$ 
\\
Boltzmann's constant\index{constant!Boltzmann's} 
& $k_{B}$\index{$k_{B}$} & $1.3806488 \times 10^{-23}$ & J~K$^{-1}$
\\
                     &         & $8.6173324 \times 10^{-11}$ & MeV~K$^{-1}$ 
\\
gravitational constant\index{constant!gravitational} 
& $G$\index{$G$} & $6.67384 \times 10^{-11}$ & m$^3$~kg$^{-1}$~s$^{-2}$
\\
                       &     & $6.70837 \times 10^{-39}$ &
                       GeV$^{-2}c^{-4}$
\\
 fine structure constant\index{constant!fine structure} 
 & $\alpha=e^{2}/(\hbar c)$\index{$\alpha$}\index{$e^{2}$} &
 $1/137.035999074$ & $-$ \\
 neutron mass\index{mass!neutron} & $m_{n}$\index{$m_{n}$} & $939.565379$ & MeV~$c^{-2}$ \\
 proton mass\index{mass!proton}  & $m_{p}$\index{$m_{p}$} & $938.272046$ & MeV~$c^{-2}$ \\
 electron mass\index{mass!electron} 
& $m_{e}$\index{$m_{e}$} & $0.510998928$ & MeV~$c^{-2}$ \\
 muon mass\index{mass!muon} & $m_{\mu}$\index{$m_{\mu}$} & $105.6583715$ & MeV~$c^{-2}$ 
\\
\bottomrule
\end{tabular}
\end{center}
\end{table}

\section{Physical conditions}
\label{sec:physcon}

Predictions for the 
properties of dense matter can differ considerably depending on the
employed model, the considered constituents and interactions.
In the CompOSE database, the equation of state
is considered to describe dense matter in thermodynamic 
equilibrium\index{equilibrium!thermodynamic}
that is defined as follows.
It is assumed that all the constituents are in chemical
equilibrium\index{equilibrium!chemical} regardless 
of the time scales\index{time scale} 
and reaction rates\index{reaction!rate}
for the relevant conversion reactions mediated by strong and
electromagnetic interactions. This condition leads to
relations between the 
chemical potentials\index{potential!chemical} 
of all particles.

In contrast, an equilibrium with
respect to weak\index{equilibrium!weak} interaction reactions, in particular
$\beta$-equilibrium\index{equilibrium!$\beta$}, is not supposed in
general. Similarly, 
the chemical potentials of the massive leptons\index{lepton} of all generations,
i.e.\ electrons\index{electron} and muons\index{muon}
(tauons\index{tauon} are irrelevant for the considered conditions)
cannot assumed to be equal.
There is an option available such that EoS tables can be provided
in particular cases which also take
$\beta$-equilibrium into account, reducing the number of independent
parameters. Assumptions on the relation between the electron and muon
chemical potentials are discussed in the description of each model
separately.

For EoS models with
strangeness\index{strangeness}\index{quark!strange} 
bearing particles,
e.g.\ hyperons\index{hyperon} or kaons\index{kaon}, 
it is assumed
that the strangeness chemical
potential\index{potential!chemical!strangeness} 
vanishes. This means that we assume the strangeness changing weak
interactions to be in equilibrium. Note that this is not the case in
heavy-ion collisions\index{collision!heavy-ion}, 
where, on the contrary strangeness is conserved,
i.e. there is no net strangeness.

Except for the tables of pure hadronic\index{matter!hadronic} 
or quark matter\index{matter!quark} without massive
leptons, also local charge neutrality\index{charge!neutrality} 
is assumed to hold.
Neutrinos\index{neutrino} and their contribution to thermodynamic
properties are never included in the present tables since they are
usually treated independently from the EoS in astrophysical
simulations because a thermodynamic
equilibrium\index{equilibrium!thermodynamic} 
can not be assumed in general.

Photons\index{photon} are usually included in equations of state for astrophysical
applications. Their treatment is discussed in section \ref{sec:photons}.

\section{Particle number densities and particle fractions}

Particle number densities\index{density!number} 
of all particles $i$ are given by
\begin{equation}
 n_{i}\index{$n_{i}$} = \frac{N_{i}}{V} \qquad [\mbox{fm}^{-3}]
\end{equation}
where $N_{i}$\index{$N_{i}$} [dimensionless] is the particle
number\index{particle!number} 
inside the volume\index{volume} 
$V$\index{$V$} [fm$^{3}$].
Note that for particles with half-integer spin\index{spin!half-integer},
$n_{i}$ represents the net particle density\index{density!net}, i.e.\
it is the difference between the particle and antiparticle\index{antiparticle}
density. E.g., for electrons we have $n_{e} = n_{e^{-}}-n_{e^{+}}$.
For particles with integer spin\index{spin!integer}, e.g.\ 
mesons\index{meson}, 
particle and antiparticle\index{antiparticle} 
densities\index{density!meson}\index{density!antiparticle} 
are distinguished and given separately.

>From the individual particle number densities several
new {\it composite} number densities\index{density!number!composite} 
can be deduced that are convenient to
characterize the state of the system.
The baryon number density $n_{b}$\index{density!number!baryon} 
and the total number of baryons $N_{b}$\index{number!baryon} 
[dimensionless] are given by
\begin{equation}
 n_{b}\index{$n_{b}$} = \frac{N_{b}}{V}\index{$N_{b}$} =
\sum_{i} B_{i} n_{i} \qquad [\mbox{fm}^{-3}]
\end{equation}
with the baryon number\index{baryon!number} 
$B_{i}$\index{$B_{i}$} [dimensionless] for a particle $i$. 
The baryon number of a nucleus
$i$ is just the mass number $A_{i}$\index{$A_{i}$}
[dimensionless] and for a 
quark\index{quark!baryon number} $i$ one has $B_{i}= 1/3$.
{\bf Warning:} In many astrophysical applications, a mass
density\index{density!mass} $\varrho = m \, n_{b}$\index{$\varrho$} 
is introduced as a parameter with
$m$ representing, e.g.\ the neutron mass\index{mass!neutron} or the
atomic mass unit\index{atomic mass unit}.
But, 
$\varrho$\index{$\varrho$} is not identical
to the total mass\index{density!mass!total} 
($=$ total internal energy\index{energy!internal!total}) density. 
Hence, $\varrho V$ is not a conserved quantity.

The strangeness number density $n_{s}$\index{density!number!strangeness}
and total strangeness number $N_{s}$\index{number!strangeness}
\begin{equation}
 n_{s}\index{$n_{s}$} = \frac{N_{s}}{V}\index{$N_{s}$} =
\sum_{i} S_{i} n_{i} \qquad [\mbox{fm}^{-3}]
\end{equation}
and the lepton number densities $n_{le}$, $n_{l\mu}$\index{density!number!lepton}
and number of leptons $N_{le}$, $N_{l\mu}$\index{lepton!number} 
\begin{eqnarray}
 n_{le}\index{$n_{le}$} = \frac{N_{le}}{V}\index{$N_{le}$}
& = & \sum_{i} L^{e}_{i} n_{i} \qquad [\mbox{fm}^{-3}]
 \\
 n_{l\mu}\index{$n_{l\mu}$} = \frac{N_{l\mu}}{V}\index{$N_{l\mu}$}
& = & \sum_{i} L^{\mu}_{i} n_{i} \qquad [\mbox{fm}^{-3}]
\end{eqnarray}
with strangenes numbers\index{strangeness!number}
$S_{i}$\index{$S_{i}$} [dimensionless]
and lepton numbers
 $L^{e}_{i}$\index{$L^{e}_{i}$},
 $L^{\mu}_{i}$\index{$L^{\mu}_{i}$} 
[dimensionless], respectively, 
are defined similar to the baryon number density. 
Since neutrinos are not included in the CompOSE EoS, $n_{le} =
n_{e}$\index{$n_{e}$} and $n_{l\mu}=n_{\mu}$\index{$n_{\mu}$}. 
We also consider the charge density of strongly
  interacting particles $n_{q}$\index{density!charge}
and corresponding charge number $N_{q}$\index{number!charge}
\begin{equation}
 n_{q}\index{$n_{q}$} = \frac{N_{q}}{V}\index{$N_{q}$} 
= \sum_{i}{}^{\prime}\index{$\sum_{i}{}^{\prime}$} 
 Q_{i} n_{i} \qquad [\mbox{fm}^{-3}]
\end{equation}
with charge 
numbers\index{charge!number} $Q_{i}$\index{$Q_{i}$}
[dimensionless]. The prime at the
sum symbol indicates that the summation runs over all particles
(including quarks) except leptons. 
For a nucleus\index{nucleus} 
${}^{A_{i}}Z_{i}$, the baryon number $B_{i}$ and the charge number
$Q_{i}$ are simply given by the mass number $A_{i}$\index{$A_{i}$} 
and atomic number $Z_{i}$\index{$Z_{i}$}, respectively.
Table \ref{tab:partindex} summarizes baryon, strangeness, charge 
and lepton numbers of the most relevant particles
that appear in the definition of the composite number densities.

Corresponding to the particle number densities $n_{i}$
the particle number fractions\index{fraction!particle number}
\begin{equation}
\label{eq:Ydef}
 Y_{i}\index{$Y_{i}$} = \frac{n_{i}}{n_{b}} \qquad [\mbox{dimensionless}]
\end{equation}
are defined. Due to this definition with $n_{b}$ we have the
normalization condition
\begin{equation}
 \sum_{i}{} B_{i} Y_{i} = 1
\end{equation}
with a summation over all particles. Note that $Y_{i}$ is not
necessarily
identical to the particle mass number fraction\index{fraction!particle
mass}
\begin{equation}
 X_{i} = B_{i}Y_{i}\index{$X_{i}$} \qquad [\mbox{dimensionless}]
\end{equation}
that are often introduced. We prefer to use $Y_{i}$ and not $X_{i}$ 
since the latter quantity is zero
for all particles with baryon number $B_{i}=0$, e.g.\ mesons.

Because of the imposed physical conditions (see \ref{sec:physcon}), 
the state of the system
is uniquely characterized by the three quantities
temperature $T$\index{$T$} [MeV], baryon number density
$n_{b}$\index{$n_{b}$} [fm$^{-3}$]
and charge density of strongly interaction particles
$n_{q}$\index{$n_{q}$} [fm$^{-3}$]. 
Charge neutrality\index{charge!neutrality} implies 
$n_{q} = n_{le}+n_{l\mu}$\index{$n_{le}$}\index{$n_{l\mu}$} and the strangeness number 
density\index{density!number!strangeness} $n_{s}$\index{$n_{s}$} is fixed by
the condition $\mu_{s}=0$\index{$\mu_{s}$}. 
Instead of $n_{q}$ it is more convenient
to use the charge fraction
of strongly interacting particles\index{fraction!charge}
\begin{equation}
 Y_{q}\index{$Y_{q}$} = \frac{n_{q}}{n_{b}} \qquad [\mbox{dimensionless}]
\end{equation}
as the third independent quantity.
The choice of $Y_{q}$ instead of the electron fraction\index{fraction!electron}
\begin{equation}
 Y_{e}\index{$Y_{e}$} = \frac{n_{e}}{n_{b}} \qquad [\mbox{dimensionless}]
\end{equation}
as a parameter is motivated by the
following facts. 
In pure hadronic (quark) equations of state (i.e.\ without leptons)
only $Y_{q}$ and not $Y_{e}$ is defined.
In EoS models with the condition of charge neutrality\index{charge!neutrality}
that contain electrons as the only considered charged lepton, 
the charge fraction of strongly interacting particles
$Y_{q}$ is identical to the electronic charge fraction $Y_{e}$.
In models with electrons and muons, charge neutrality requires
\begin{equation}
 Y_{q} = Y_{e}+Y_{\mu} = Y_{l}\index{$Y_{l}$}
\end{equation}
with the muon fraction
\begin{equation}
 Y_{\mu}\index{$Y_{\mu}$} 
= \frac{n_{\mu}}{n_{b}} \qquad [\mbox{dimensionless}] 
\end{equation}
and the total lepton fraction $Y_{l}$\index{fraction!lepton!total} [dimensionless].
In this case, the balance between the electron and muon densities
depends on the assumed relation of the electron and muon chemical
potentials.


\begin{table}[ht]
\begin{center}
\caption{\label{tab:partindex}%
Baryon\index{baryon!number} ($B_{i}$\index{$B_{i}$}), 
strangeness\index{strangeness!number} ($S_{i}$\index{$S_{i}$}), 
charge\index{charge!number} ($Q_{i}$\index{$Q_{i}$})
and lepton\index{lepton!number} ($L^{e}_{i}$\index{$L^{e}_{i}$},
$L^{\mu}_{i}$\index{$L^{\mu}_{i}$})
numbers and indices\index{particle!index} $I_{i}\index{$I_{i}$}$ 
of the most relevant particles $i$
in dense matter.}
{\small 
\begin{tabular}{llrrrrrr}
\toprule
particle class & symbol of particle $i$
 & $B_{i}$ & $S_{i}$ & $Q_{i}$ & $L^{e}_{i}$ & $L^{\mu}_{i}$ & 
 particle index $I_{i}$ \\
\toprule
leptons & $e^{-}$       & $0$ & $0$ & $-1$ & 1 & 0 & 0 \\
        & $\mu^{-}$     & $0$ & $0$ & $-1$ & 0 & 1 & 1 \\
\midrule
 nuclei ($A > 1$) & ${}^{Z}A$ & $A$\index{$A$}   
 & $0$   & $Z$\index{$Z$} & $0$ & $0$ & $1000 \cdot A+Z$   \\
\midrule
baryons & $n$\index{$n$}          & $1$ & $0$  & $0$  & 0 & 0 & 10  \\
        & $p$\index{$p$}          & $1$ & $0$  & $+1$ & 0 & 0 & 11  \\
        & $\Delta^{-}$\index{$\Delta^{-}$}  & $1$ & $0$  & $-1$ & 0 & 0 & 20 \\
        & $\Delta^{0}$\index{$\Delta^{0}$}  & $1$ & $0$  & $0$  & 0 & 0 & 21 \\
        & $\Delta^{+}$\index{$\Delta^{+}$}  & $1$ & $0$  & $+1$ & 0 & 0 & 22 \\
        & $\Delta^{++}$\index{$\Delta^{++}$} & $1$ & $0$  & $+2$ & 0 & 0 & 23 \\
        & $\Lambda$\index{$\Lambda$}    & $1$ & $-1$  & $0$    & 0 & 0 & 100  \\
        & $\Sigma^{-}$\index{$\Sigma^{-}$}  & $1$ & $-1$  & $-1$   & 0 & 0 & 110  \\
        & $\Sigma^{0}$\index{$\Sigma^{0}$}  & $1$ & $-1$  & $0$    & 0 & 0 & 111  \\
        & $\Sigma^{+}$\index{$\Sigma^{+}$}  & $1$ & $-1$  & $+1$   & 0 & 0 & 112  \\
        & $\Xi^{-}$\index{$\Xi^{-}$}     & $1$ & $-2$  & $-1$   & 0 & 0 & 120  \\
        & $\Xi^{0}$\index{$\Xi^{0}$}     & $1$ & $-2$  & $0$    & 0 & 0 & 121  \\
\midrule
mesons  & $\omega$\index{$\omega$}      & $0$ & $0$ & $0$ & 0 & 0 & 200 \\
        & $\sigma$\index{$\sigma$}      & $0$ & $0$ & $0$ & 0 & 0 & 210 \\
        & $\eta$\index{$\eta$}        & $0$ & $0$ & $0$ & 0 & 0 & 220 \\
        & $\eta^{\prime}$\index{$\eta^{\prime}$} & $0$ & $0$ & $0$ & 0 & 0 &230 \\
        & $\rho^{-}$\index{$\rho^{-}$}  & $0$ & $0$ & $-1$ & 0 & 0 & 300 \\
        & $\rho^{0}$\index{$\rho^{0}$}  & $0$ & $0$ & $0$  & 0 & 0 & 301 \\
        & $\rho^{+}$\index{$\rho^{+}$}  & $0$ & $0$ & $+1$ & 0 & 0 & 302 \\
        & $\delta^{-}$\index{$\delta^{-}$} & $0$ & $0$ & $-1$ & 0 & 0 & 310 \\
        & $\delta^{0}$\index{$\delta^{0}$} & $0$ & $0$ & $0$ & 0 & 0 & 311 \\
        & $\delta^{+}$\index{$\delta^{+}$} & $0$ & $0$ & $+1$ & 0 & 0 & 312 \\
        & $\pi^{-}$\index{$\pi^{-}$}    & $0$ & $0$ & $-1$ & 0 & 0 & 320 \\
        & $\pi^{0}$\index{$\pi^{0}$}    & $0$ & $0$ & $0$  & 0 & 0 & 321 \\
        & $\pi^{+}$\index{$\pi^{+}$}    & $0$ & $0$ & $+1$ & 0 & 0 & 322 \\
        & $\phi$\index{$\phi$}       & $0$ & $0$ & $0$ & 0 & 0 & 400 \\
        & $\sigma_{s}$\index{$\sigma_{s}$}  & $0$ & $0$ & $0$ & 0 & 0 & 410 \\
        & $K^{-}$\index{$K^{-}$}       & $0$ & $-1$ & $-1$ & 0 & 0 & 420 \\
        & $K^{0}$\index{$K^{0}$}       & $0$ & $+1$ & $0$ & 0 & 0 & 421 \\
        & $\bar{K}^{0}$\index{$\bar{K}^{0}$} & $0$ & $-1$ & $0$ & 0 & 0 & 422 \\
        & $K^{+}$\index{$K^{+}$}       & $0$ & $+1$ & $+1$ & 0 & 0 & 423 \\
\midrule
quarks  & $u$\index{$u$}       & $1/3$ & $0$   & $+2/3$ & 0 & 0 & 500 \\
        & $d$\index{$d$}       & $1/3$ & $0$   & $-1/3$ & 0 & 0 & 501 \\
        & $s$\index{$s$}       & $1/3$ & $-1$  & $-1/3$ & 0 & 0 & 502 \\
\midrule
photon & $\gamma$\index{$\gamma$}   & $0$   & $0$   & $0$    & 0 & 0 & 600 \\
\bottomrule
\end{tabular}
}
\end{center}
\end{table}

\begin{table}[ht]
\begin{center}
\caption{\label{tab:corrindex}%
Baryon\index{baryon!number} ($B_{i}$\index{$B_{i}$}), 
strangeness\index{strangeness!number} ($S_{i}$\index{$S_{i}$}) and 
charge\index{charge!number} ($Q_{i}$\index{$Q_{i}$})
numbers and indices\index{particle!index} $I_{i}\index{$I_{i}$}$ 
of the most relevant two-particle correlations $i$
in dense matter. The lepton numbers $L^{e}_{i}$\index{$L^{e}_{i}$} 
and $L^{\mu}_{i}$\index{$L^{\mu}_{i}$} are
always zero for these correlations.}
{\small
\begin{tabular}{lccrrrr}
\toprule
correlation class & particles & channel 
 & $B_{i}$\index{$B_{i}$} & $S_{i}$\index{$S_{i}$} & 
 $Q_{i}$\index{$Q_{i}$} & index $I_{i}$ of correlation\\
\toprule
two-body & $nn$ & ${}^{1}S_{0}$\index{${}^{1}S_{0}$} & $2$ & $0$ & $0$ & 700 \\
         & $np$ & ${}^{1}S_{0}$ & $2$ & $0$ & $1$ & 701 \\
         & $pp$ & ${}^{1}S_{0}$ & $2$ & $0$ & $2$ & 702 \\
         & $np$ & ${}^{3}S_{1}$\index{${}^{3}S_{1}$} & $2$ & $0$ & $1$ & 703 \\
\bottomrule
\end{tabular}
}
\end{center}
\end{table}

\section{Particle indexing}

The composition\index{composition} 
of dense matter can rapidly change with 
temperature\index{temperature} $T$ [MeV],
baryon number density\index{density!number!baryon} 
$n_{b}$ [fm${}^{-3}$] and the charge
fraction of strongly interacting particles\index{fraction!charge} $Y_{q}$
[dimensionless]. We introduce
an indexing\index{indexing!particle} scheme for the 
particles\index{particle!index} that allows to identify them
uniquely in order to store only the most
abundant particles in the EoS tables. In table \ref{tab:partindex}
an overview of indices for most of the relevant particles is
presented. Indices of missing particles can be added on request.

In addition to the various particles that can appear in dense matter,
there is the possibility of strong two- or even three-particle
correlations\index{correlation} such as pairing\index{pairing} of nucleons or quarks 
and phenomena such as superfluidity can arise. Hence, an indexing
scheme for identifying these channels is introduced, too. In table
\ref{tab:corrindex} the notation is given for these correlations.  
Again, indices of missing particle correlations can be added on request.

\section{Thermodynamic potentials and basic quantities}
\label{sec:thpot}

All thermodynamic properties\index{property!thermodynamic} 
of a system are completely determined
if a thermodynamic potential is known as a function of its 
natural variables.
In the general case, this can be formulated as follows.
The properties can be derived
from the thermodynamic potential\index{potential!thermodynamic} 
$\Xi = \Xi(x_{i},\xi_j)$\index{$\Xi$}
depending on $n$ natural variables\index{variable!natural} 
$x_{i}$, $i=1,\dots,n_{1}$\index{$x_{i}$} and $\xi_{j}$, $j = 1\dots,
n_{2}$\index{$\xi_{j}$} with $n_{1} + n_{2} = n$.
The quantities $x_{i}$ and $\xi_{i}$ represent extensive and intensive
variables\index{variable!intensive}\index{variable!extensive}, respectively.
The relations
\begin{equation}
 \xi_{i} = \left. \frac{\partial \Xi}{\partial x_{i}} \right|_{x_{k}, k \neq i; \xi_{j}}
 = \xi_{i}(x_{i},\xi_{j})
\end{equation}
and
\begin{equation}
 x_{j} =  -\left. \frac{\partial \Xi}{\partial \xi_{j}} \right|_{\xi_{k}, k \neq j; x_{i}}
 = x_{j}(x_{i},\xi_{j})
\end{equation}
are the thermodynamic equations of state\index{equation of state} and
define the variables\index{variable!conjugate} 
$\xi_{i}$ that are conjugate to $x_{i}$
as first partial derivatives of $\Xi$ and vice versa.
Due to Euler's theorem\index{theorem!Euler's} 
on homogeneous functions\index{function!homogeneous}, the
thermodynamic potential is given by the sum
\begin{equation}
\label{eq:xix}
 \Xi(x_{i},\xi_{j}) = \sum_{i=1}^{n_{1}} \xi_{i} x_{i}
\end{equation}
running over all 
$i=1,\dots,n_{1}$ intensive variables\index{variable!intensive}. 
Thus, the knowledge of all relevant equations of
state or first derivatives is sufficient to recover the thermodynamic
potential $\Xi$ completely.
For the mixed second partial derivatives of a thermodynamic potential $\Xi$
the result is independent of the order of differentiation
and the Maxwell relations\index{relation!Maxwell}
\begin{equation}
 \left. \frac{\partial \xi_{l}}{\partial x_{k}} 
 \right|_{x_{j}, j\neq k} 
 = \left. \frac{\partial \xi_{k}}{\partial x_{l}} 
 \right|_{x_{j}, j \neq l} 
\end{equation}
are obtained. For a more detailed discussion of these and further
aspects we refer the reader to standard text books on thermodynamics.

In most models for the EoS of dense matter, 
the temperature\index{temperature} $T$\index{$T$} [MeV], 
the volume\index{volume} $V$\index{$V$}
[fm$^{3}$] and the individual 
particle numbers\index{particle!number} $N_{i}$ \index{$N_{i}$} [dimensionless]
are selected as natural variables. This case corresponds to the (Helmholtz) free
energy\index{energy!free}\index{energy!Helmholtz free} 
$F=F(T,V,N_{i})$\index{$F$}
[MeV] as the relevant thermodynamic potential that contains all
information. Note that we assume that the free energy includes contributions 
by the rest masses\index{rest mass} 
of the particles. Keeping the volume $V$ fixed,
it is convenient to define the free energy density\index{density!energy!free}
\begin{equation}
 f(T,n_{i})\index{$f$} = \frac{F}{V} \qquad [\mbox{MeV~fm}^{-3}] \: ,
\end{equation} 
and the entropy density\index{density!entropy}
\begin{equation}
 s(T,n_{i})\index{$s$} =  - \left. \frac{\partial f}{\partial
     T}\right|_{n_{i}}  \qquad [\mbox{fm}^{-3}]
\end{equation}
with the entropy\index{entropy}
\begin{equation}
 S(T,V,N_{i})\index{$S$} = V s(T,n_{i}) = - \left. \frac{\partial F}{\partial
     T}\right|_{V,N_{i}} \qquad [\mbox{dimensionless}] \: .
\end{equation}
The chemical potential\index{potential!chemical} of a particle $i$ is given by
\begin{equation}
 \mu_{i}\index{$\mu_{i}$} = \left. \frac{\partial F}{\partial N_{i}}
 \right|_{T,V,N_{j},j\neq i} =  \left. \frac{\partial f}{\partial n_{i}}
 \right|_{T,n_{j},j\neq i}
 \qquad [\mbox{MeV}] 
\end{equation}
including the rest mass $m_{i}$\index{$m_{i}$} [MeV].
The pressure\index{pressure} is obtained from
\begin{equation}
 p\index{$p$} = - \left. \frac{\partial F}{\partial V}
 \right|_{T,N_{i}} 
 =  n_{b}^{2} \left. \frac{\partial (f/n_{b})}{\partial n_{b}}
 \right|_{T,Y_{q}} 
 = \sum_{i} \mu_{i} n_{i} - f
 \qquad [\mbox{MeV~fm}^{-3}]  \: .
\end{equation}

Each of the composite densities\index{density!composite} 
$n_{b}$, $n_{s}$, $n_{le}$, $n_{l\mu}$, $n_{q}$
\index{$n_{b}$}\index{$n_{s}$}\index{$n_{le}$}\index{$n_{l\mu}$}\index{$n_{q}$}
is accompanied by a 
corresponding chemical potential,
i.e.\ we have the 
baryon number chemical potential\index{potential!chemical!baryon} 
$\mu_{b}$\index{$\mu_{b}$} [MeV],
the strangeness number chemical
potential\index{potential!chemical!strangeness} 
$\mu_{s}$\index{$\mu_{s}$} [MeV],
the electron lepton number chemical potential\index{potential!chemical!electron} 
$\mu_{le}$\index{$\mu_{le}$} [MeV],
the muon lepton number chemical potential\index{potential!chemical!muon} 
$\mu_{l\mu}$\index{$\mu_{l\mu}$} [MeV],
and the charge chemical
potential\index{potential!chemical!charge} 
$\mu_{q}$\index{$\mu_{q}$} [MeV].
The chemical potential
of a particle $i$ is then given by
\begin{equation}
  \mu_{i}\index{$\mu_{i}$} = B_{i} \mu_{b} + Q_{i} \mu_{q} + S_{i}
  \mu_{s} + L^{e}_{i} \mu_{le} + L^{\mu}_{i} \mu_{l\mu}
  \qquad [\mbox{MeV}] \: ,
\end{equation}
e.g.\ $\mu_{n} = \mu_{b}$, $\mu_{p}=\mu_{b}+\mu_{q}$ and $\mu_{e} = \mu_{le}-\mu_{q}$.
Note again that we use the relativistic definition of chemical potentials
including rest masses\index{rest mass}.
It has to be mentioned that $n_{b}$, $n_{s}$, $n_{le}$ and $n_{l\mu}$
are conjugate to $\mu_{b}$, $\mu_{s}$, $\mu_{le}$ and $\mu_{l\mu}$,
respectively. However, this is not true for $n_{q}$ and $\mu_{q}$ if
leptons are included in the EoS.
In general, it is assumed that $\mu_{s}=0$ since
strangeness changing weak interaction reactions are in equilibrium.
Nevertheless, the strangeness density $n_{s}$ can
be non-zero if 
particles with strangeness are considered in the EoS.

In addition to the free energy density, several other thermodynamic
potentials can be defined by applying Legendre
transformations\index{transformation!Legendre}, e.g.\ 
the internal energy\index{energy!internal}
\begin{equation}
 E\index{$E$} = E(S,V,N_{i}) = F+TS  \qquad [\mbox{MeV}] \: ,
\end{equation}
the free enthalpy\index{enthalpy!free} (Gibbs potential\index{potential!Gibbs})
\begin{equation}
 G\index{$G$} 
 = G(T,p,N_{i}) = F+pV = \sum_{i}\mu_{i}N_{i}   \qquad [\mbox{MeV}] \: ,
\end{equation}
the enthalpy\index{enthalpy}
\begin{equation}
 H\index{$H$} = H(S,p,N_{i}) = E+pV   \qquad [\mbox{MeV}] \: ,
\end{equation}
and the grand canonical potential\index{potential!grand canonical}
\begin{equation}
 \Omega\index{$\Omega$} = \Omega(T,V,\mu_{i}) = F-\sum_{i}\mu_{i}N_{i} = -pV 
  \qquad [\mbox{MeV}] 
\end{equation}
with the corresponding densities $e=E/V$\index{$e$}, 
$g=G/V$\index{$g$}, $h=H/V$\index{$h$} and 
$\omega = \Omega/V$\index{$\omega$} [MeV~fm${}^{-3}$], respectively.

It is convenient to define the free energy per baryon
\begin{equation}
 \mathcal{F}\index{$\mathcal{F}$} 
 = \frac{F}{N_{b}} = \frac{f}{n_{b}} \qquad [\mbox{MeV}] \: ,
\end{equation}
the internal energy per baryon
\begin{equation}
 \mathcal{E}\index{$\mathcal{E}$} 
 = \frac{E}{N_{b}} = \frac{e}{n_{b}} \qquad [\mbox{MeV}] \: ,
\end{equation}
the enthalpy per baryon
\begin{equation}
 \mathcal{H}\index{$\mathcal{H}$}
 = \frac{H}{N_{b}} = \frac{h}{n_{b}} \qquad [\mbox{MeV}] \: ,
\end{equation}
and 
the free enthalpy per baryon
\begin{equation}
 \mathcal{G}\index{$\mathcal{G}$}
 = \frac{G}{N_{b}} = \frac{g}{n_{b}} \qquad [\mbox{MeV}] 
\end{equation}
by dividing the corresponding thermodynamic potential by the 
total number of baryons
\begin{equation}
 N_{b}\index{$N_{b}$} = n_{b} V \qquad [\mbox{dimensionless}] \: .
\end{equation}

\section{Thermodynamic coefficients}
\label{sec:thermo_coeff}

\index{coefficient!thermodynamic}
In many applications tabulated values of 
the thermodynamic potentials and their first
derivatives are not sufficient and additional quantities that depend
on second derivatives of the thermodynamic potentials are needed.
In general, the relevant thermodynamic potential can depend on
a large number of independent variables such as temperature, volume and  particle
densities or chemical potentials. In applications, however, this
number often reduces due to physical constraints. E.g.,
the condition of chemical equilibrium between certain particle species
reduces the number of independent chemical potentials. In a simular
way, the condition of charge neutrality relates the particle numbers
of charged constituents. In the following, we consider only systems
that are
completely determined by the temperature $T$, the volume $V$, the
total number of baryons $N_{b}= n_{b} V $ and the charge number 
$N_{q} = Y_{q} N_{b} = Y_{q} n_{b} V$, or in the case of an
equation of state where leptons are present the number of charged leptons,
which are equal to $N_q$ due to charge neutrality. Then
we have with the free energy per baryon $\mathcal{F}(T,n_{b},Y_{q})$ [MeV]
the specific heat capacity at constant volume\index{heat capacity}
\begin{equation}
 c_{V}\index{$c_{V}$} = \frac{T}{N_{b}} \left. \frac{dS}{dT} \right|_{V,N_{b},N_{q}}
 = -T \left. \frac{\partial^{2} \mathcal{F}}{\partial
     T^{2}}\right|_{n_{b},Y_{q}} \qquad [\mbox{dimensionless}] \: ,
\end{equation}
the tension coefficient at constant volume\index{coefficient!tension}
\begin{equation}
 \beta_{V}\index{$\beta_{V}$} =  \left. \frac{dp}{dT} \right|_{V,N_{b},N_{q}}
 = \left. \frac{dS}{dV} \right|_{T,N_{b},N_{q}}
 = n_{b}^{2} \left. \frac{\partial^{2} \mathcal{F}}{\partial T \partial
     n_{b}}\right|_{Y_{q}} \qquad [\mbox{fm}^{-3}] \: ,
\end{equation}
the isothermal compressibility\index{compressibility!isothermal}
\begin{eqnarray}
\label{eq:kappat}
 \kappa_{T}\index{$\kappa_{T}$}
 & = & -\frac{1}{V} \left. \frac{dV}{dp} \right|_{T,N_{b},N_{q}}
 =   \left( n_{b} \left. \frac{\partial p}{\partial n_{b}}
 \right|_{T,Y_{q}} \right)^{-1}
 \\ \nonumber & = & 
   \left( n_{b}^{2} \left. \frac{\partial^{2} (\mathcal{F}n_{b})}{\partial n_{b}^{2}}
     \right|_{T,Y_{q}} \right)^{-1}
\qquad [\mbox{MeV}^{-1}\mbox{fm}^{3}] \: ,
\end{eqnarray}
the expansion coefficient at constant pressure\index{coefficient!expansion}
\begin{equation}
 \alpha_{p}\index{$\alpha_{p}$} 
 = \frac{1}{V} \left. \frac{dV}{dT} \right|_{p,N_{b},N_{q}}
 = 
  \kappa_{T} \beta_{V}  \qquad [\mbox{MeV}^{-1}] \: ,
\end{equation}
the specific heat capacity at constant pressure\index{heat capacity}
\begin{equation}
 c_{p}\index{$c_{p}$} = \frac{T}{N_{b}} \left. \frac{dS}{dT}
 \right|_{p,N_{b},N_{q}}
 = c_{V} 
 + \frac{T}{n_{b}} \alpha_{p} \beta_{V}
 \qquad [\mbox{dimensionless}] \: ,
\end{equation}
the adiabatic index\index{index!adiabatic}
\begin{equation}
 \Gamma\index{$\Gamma$} = \frac{c_{p}}{c_{V}} \qquad [\mbox{dimensionless}] 
\end{equation}
and the adiabatic compressibility\index{compressibility!adiabatic}
\begin{equation}
 \kappa_{S}\index{$\kappa_{S}$} 
 = - \frac{1}{V} \left. \frac{dV}{dp} \right|_{S,N_{b},N_{q}}
 = \frac{\kappa_{T}}{\Gamma}  \qquad [\mbox{MeV}^{-1}\mbox{fm}^{3}] \: .
\end{equation}
The square of the 
speed of sound\index{speed of sound} (isoscalar longitudinal
compression wave)
in the medium for a one-fluid flow is given by the relativistic definition
\begin{equation}
 c_{s}^{2}\index{$c_{s}$} 
 =  \left. \frac{dp}{de} \right|_{S,N_{b},N_{q}} 
 = \frac{1}{h\kappa_{S}} 
  \qquad [\mbox{dimensionless}]
\end{equation}
with the enthalpy density $h=e+p$ [MeV~fm$^{-3}$].
Alternatively, the expression
\begin{equation}
c_{s}^{2} =  
  \left. \frac{n_{b}}{h} \frac{\partial p}{\partial n_{b}}
   \right|_{Y_{q},e} 
 +  \left. \frac{\partial p}{\partial e} 
 \right|_{n_{b},n_{q}} 
  \qquad [\mbox{dimensionless}]
\end{equation}
can be used.
All of the above quantities can be calculated with the help
of the second derivatives of $\mathcal{F}$ or $f$ with respect to the
parameters $T$, $n_{b}$, and $n_{q}$.

\section{Microscopic quantities}

Most EoS models are based on more or less microscopic models that
can give information on microscopic particle properties
in addition to the thermodynamic and compositional
quantities. Only few 
models are just simple parametrizations of thermodynamic quantities
without recurrence to the underlying microphysics. The
CompOSE database allows to store these microscopic values, too, 
such that they can be used in applications. 
In the present version, 
we consider a small set of quantities that are discussed in the
following. It can be easily extended if required.

In particular, we consider the effective masses of the particles where
one has to distinguish between different definitions.
The effective Landau mass\index{effective mass!Landau}
\begin{equation}
 m^{L}_{i}\index{$m^{L}_{i}$} \qquad [\mbox{MeV}]
\end{equation}
of a particle $i$ is related to the single-particle density of
states\index{density of states} and defined as
\begin{equation}
\frac{1}{m^{L}_{i}} = \frac{1}{p^{F}_{i}} \left. \frac{d E_{i}}{d
    p}\right|_{p = p_{i}^{F}}
 \: ,
\end{equation}
where $p^{F}_{i}$ denotes the respective Fermi-momentum and $E_{i}$ is the
single-particle energy. In non-relativistic Skyrme-Hartree-Fock models it
appears in the 
kinetic energy\index{energy!kinetic} contribution 
\begin{equation}
 T_{i}\index{$T_{i}$} = \frac{p_{i}^{2}}{2m^{L}_{i}} \qquad [\mbox{MeV}]
\end{equation}
to the total single-particle energy.
In addition to the kinetic single-particle energy $T_{i}$, a single-particle
potential\index{potential!single-particle}
\begin{equation}
 U_{i}\index{$U_{i}$} \qquad [\mbox{MeV}]
\end{equation}
contributes to the single-particle energy\index{energy!single-particle}
\begin{equation}
E_{i}\index{$E_{i}$} = T_{i} + U_{i} + m_{i}
\end{equation}
in nonrelativistic models. 
The effective Dirac mass\index{effective mass!Dirac}
\begin{equation}
 m^{D}_{i}\index{$m^{D}_{i}$} = m_{i} - S_{i} \qquad [\mbox{MeV}]
\end{equation}
is found in the single-particle Hamiltonian of relativistic models,
such as relativistic mean-field approaches. This quantity does not
reflect directly the density of single-particle states. At zero
temperature the corresponding effective Landau mass can be obtained
from
\begin{equation}
 m^{L}_{i} = \sqrt{(m^{D}_{i})^{2}+(p^{F}_{i})^{2}} \: .
\end{equation}
The effective Dirac mass depends on the
scalar self-energy\index{self-energy!scalar} 
\begin{equation}
 S_{i}\index{$S_{i}$} = \Sigma_{i}\index{$\Sigma_{i}$} \qquad [\mbox{MeV}]
\end{equation}
that enters the relativistic single-particle Hamiltonian together with the
vector self-energy\index{self-energy!vector}
\begin{equation}
 V_{i}\index{$V_{i}$} = \Sigma^{0}_{i}\qquad [\mbox{MeV}] 
\end{equation}
that is the time-component of the general four-vector self-energy
$\Sigma^{\mu}_{i}$.\index{$\Sigma^{\mu}_{i}$}

In certain parameter regions of an EoS, the phenomena of
superconductivity\index{superconductivity} 
or superfluidity\index{superfluidity} can be found. In this case
it is worthwhile to know the size of the pairing gaps\index{pairing!gap}
\begin{equation}
 \Delta_{i}\index{$\Delta_{i}$} \qquad [\mbox{MeV}] \: ,
\end{equation}
e.g. in the $i=nn({}^{1}S_{0})$ channel in case of 
neutron pairing\index{pairing!neutron}.

\section{Contribution of photons}
\label{sec:photons}

If photons\index{photon} are considered in an EoS, they are treated as a gas of
massless bosons with simple analytical expressions for the relevant basic
thermodynamic quantities. The contribution of the photon can be
simply added to the
thermodynamic quantities of the other constituents. We have
the photon free energy density
\begin{equation}
 f_{\gamma}(T)\index{$f_{\gamma}$} 
 = -\frac{\pi^{2}}{45} T^{4} \qquad [\mbox{fm}^{-3}]
\end{equation}
the photon entropy density
\begin{equation}
 s_{\gamma}(T)\index{$s_{\gamma}$} 
 = \frac{4\pi^{2}}{45} T^{3} \qquad [\mbox{fm}^{-3}]
\end{equation}
and the photon pressure
\begin{equation}
 p_{\gamma}(T)\index{$p_{\gamma}$} 
 = \frac{\pi^{2}}{45} T^{4} \qquad [\mbox{MeV~fm}^{-3}]
\end{equation}
depending only on the temperature.  The chemical potential of the photon
vanishes
\begin{equation}
 \mu_{\gamma}\index{$\mu_{\gamma}$} = 0 \qquad [\mbox{MeV}]
\end{equation}
since it is its own antiparticle.
>From these quantities, all other relevant photon thermodynamic
quantities can be derived.

\part{For Contributors}
\label{part:contrib}
The success of the CompOSE data base depends on the support
of nuclear physicists providing tables with their favourite EoS.
Some well known EoS used extensively in astrophysical applications
are already incorporated in the CompOSE data base and are ready for use. 
However, a larger collection of EoS from different models
is highly desirable.
In order to be suitable for a simple usage, contributors should follow
the guidelines specified in this part of the manual.

\chapter{How to prepare EoS  tables}

\index{table!EoS}
In general, tables with EoS data contain a wealth of information on the
thermodynamic properties, the composition of dense matter and on the
microphysical properties of the constituents.
In order to minimize the memory size, only the essential
thermodynamic quantities should be stored in the tables that are
used in the CompOSE data base. These quantities are selected such that
they are sufficient to calculate a particular thermodynamic
potential, here the free energy\index{energy!free} $F$\index{$F$} [MeV], 
and its first derivatives with respect to the
parameters that correspond to the natural
variables\index{variable!natural} 
of $F$, i.e.\
temperature\index{temperature} $T$\index{$T$} [MeV], 
volume\index{volume} $V$\index{$V$} [fm${}^{3}$] 
and particle numbers\index{particle!number} 
$N_{i}$\index{$N_{i}$} [dimensionless].
However, it is worthwhile to have redundant information in order
to check the consistency of the EoS tables or to simplify the
determination of additional quantities. 
Further thermodynamic quantities relevant for astrophysical
applications can, if not directly available, be generated with the help of
thermodynamic identities and, for quantities depending on
higher derivatives, by using appropriate
interpolation\index{interpolation} schemes.
The organization of the data 
concerning composition etc.\ depends on the 
particular EoS. Thus it is required that three separate tables are
provided with unique formats\index{table!format} that  contain the 
thermodynamic quantities, the 
information on the composition of matter (depending
on the choice of the constituents of the model) and the microscopic
properties of the particles, respectively.
Details of the employed discretization\index{discretization} mesh of the
parameters have to be supplied in additional tables.
Additional information that is not contained in the Eos tables has to
be provided by the contributor. It will appear in the data
sheet\index{data sheet} that
accompanies each EoS and is available on the CompOSE web
site\index{web site}.

\section{Tabulation of quantities}

\subsection{Parameters and parameter ranges}
\label{ssec:para}

\index{parameter!range}
Because of the imposed physical conditions, see section \ref{sec:physcon}, 
the state of the system
is uniquely characterized by only three quantities,
see section \ref{sec:thpot}, that are used as parameters in the
EoS tables with the properties of dense matter:

\begin{enumerate}
\item {\bf temperature} $T$\index{$T$} [MeV],
\item {\bf baryon number density} $n_{b}$\index{$n_{b}$} [fm$^{-3}$],
\item {\bf charge fraction of strongly interacting particles} 
$Y_{q}$\index{$Y_{q}$} [dimensionless].
\end{enumerate}

In order to be useful in many applications, the 
parameter ranges\index{parameter!range} should be chosen as wide as possible
and the mesh spacing as fine as possible, i.e.\ the number of
grid points as large as possible. In some cases it might be
useful to adapt the resolution\index{resolution} 
of the tables, in particular for the
baryon density, to the
physical situation and occurring phenomena. E.g.\ a rapid change
in some quantities or a phase transition\index{phase!transition} can occur.
For this purpose, several tables for different blocks in density
with different resolution
can be supplied for a single EoS model.

The recommended discretization\index{discretization} scheme 
for a general purpose EoS table depends on the parameters:
\begin{enumerate}
\item {\bf temperature}\index{temperature}\index{$T$}\\
In this case, two standards are suggested:\\
a. linear mesh in $\ln T$, i.e.\
\begin{equation}
 T(i_{T}\index{$i_{T}$}) = T^{\rm ref} (f_{T})^{i_{T}-1} \quad
 f_{T} = 10^{1/M_{T}} \quad 
 i_{T}=N_{T}^{\rm min},N_{T}^{\rm min}+1, \dots, N_{T}^{\rm max} 
\end{equation}
with parameters $T^{\rm ref}\index{$T^{\rm ref}$} > 0$~MeV, 
$1 \le N_{T}^{\rm min}\index{$N_{T}^{\rm min}$} <
N_{T}^{\rm max}\index{$N_{T}^{\rm max}$}\le 101$
and $M_{T}\index{$M_{T}$}>1$ points per decade in temperature;\\
b. a mesh that is linear in $T$ at low temperatures and
linear in $\ln T$ at high temperatures, i.e.
\begin{equation}
 T(i_{T}) = T^{\rm ref} \frac{\sinh (f_{T} i_{T})}{\sinh (f_{T})} \quad
 i_{T}=N_{T}^{\rm min},N_{T}^{\rm min}+1, \dots, N_{T}^{\rm max} 
\end{equation}
 with parameters
$T^{\rm ref}>0$~MeV, $1 \le N_{T}^{\rm min} <
N_{T}^{\rm max}\le 101$ and $f_{T}>0$. The
resolution\index{resolution} 
parameter $M_{T}>0$ determines $f_{T}\index{$f_{T}$} = \ln (10)/M_{T}$;\\
In both cases the index $i_{T}=0$ is reserved for $T=0$~MeV;
\item {\bf baryon number density}\index{density!number!baryon}\index{$n_{b}$}\\
linear mesh in $\ln n_{b}$, i.e.\
\begin{equation}
 n_{b}(i_{n_{b}}\index{$i_{n_{b}}$}) = n_{b}^{\rm ref} (f_{n_{b}})^{i_{n_{b}}-1} \quad
 f_{n_{b}} = 10^{1/M_{n_{b}}} \quad
 i_{n_{b}} =N_{n_{b}}^{\rm min},N_{n_{b}}^{\rm min}+1, \dots,
 N_{n_{b}}^{\rm max} 
\end{equation} 
with parameters $n_{b}^{\rm ref}\index{$n_{b}^{\rm ref}$}>0$~fm${}^{-3}$, 
$1 \le N_{n_{b}}^{\rm  min}\index{$N_{n_{b}}^{\rm min}$} < 
N_{n_{b}}^{\rm max}\index{$N_{n_{b}}^{\rm max}$}\le 401$
and $M_{n_{b}}\index{$M_{n_{b}}$}>1$ 
points per decade in the baryon number density;
\item {\bf charge fraction of strongly interacting particles}
\index{fraction!charge}\index{$Y_{q}$}\\
linear mesh in $Y_{q}$, i.e.\
\begin{equation}
 Y_{q}(i_{Y_{q}}\index{$i_{Y_{q}}$}) = \frac{i_{Y_{q}}}{M_{Y_{q}}} \quad
 i_{Y_{q}}=N_{Y_{q}}^{\rm min},N_{Y_{q}}^{\rm min}+1, \dots,
 N_{Y_{q}}^{\rm max} 
\end{equation} 
with
parameters 
$0 \le N_{Y_{q}}^{\rm min}\index{$N_{Y_{q}}^{\rm min}$} 
< N_{Y_{q}}^{\rm max}\index{$N_{Y_{q}}^{\rm max}$}< 100$ and
$M_{Y_{q}}\index{$M_{Y_{q}}$}>N_{Y_{q}}^{\rm max}$
with a resolution in $Y_{q}$ of
$1/M_{Y_{q}}$; the indices $i_{Y_{q}}=0$ and $i_{Y_{q}}=100$
should correspond to $Y_{q}=0$ and $Y_{q}=1$, respectively.
\end{enumerate}
Each grid point\index{grid point} in the table is identified with the triple
$(i_{T},i_{n_{b}},i_{Y_{q}})$ of indices.
The recommended values for the parameters are given in section
\ref{ssec:tot}.
Actually used meshes 
for particular EoS are specified
in the online documentation for each individual EoS table.

\subsection{Thermodynamic consistency}

The equations of state included in the CompOSE data base
are required to fullfill some basic
thermodynamic consistency\index{consistency!thermodynamic}
relations. 
Of course, the consistency can only hold up to some numerical level.
Measures of the thermodynamic consistency are included in the
characterisation of the EoS tables on the web site. 

Because the free energy density is considered in the present case
as the basic thermodynamic potential,
the homogeneity\index{homogeneity} condition, cf.\ eq. (\ref{eq:xix}),
\begin{equation}
\label{eq:f_con}
 f(T,n_{b},Y_{q})\index{$f$} = 
 -p+\mu_{b} n_{b}+\mu_{q} \left( n_{q}-n_{le} - n_{l\mu}\right)  
 +\mu_{le} n_{le} +\mu_{l\mu} n_{l\mu} 
 \: .
\end{equation}
with the pressure $p$, the chemical potentials $\mu_{b}$,
$\mu_{q}$, 
$\mu_{le}$, $\mu_{l\mu}$ and corresponding densities
$n_{b}$, $n_{q}$, 
$n_{le}$, $n_{l\mu}$ should hold.
Note that 
$n_{q}$ contains only the  contribution
of strongly interacting particles
to the charge density, that 
a fixed relation between $\mu_{le}$ and $\mu_{l\mu}$ is
assumed and that $n_{le}+n_{l\mu}=n_{q} = Y_{q}n_{b}$ due to local
charge neutrality 
if leptons are considered in the equation
of state. Additionally, in eq.\ (\ref{eq:f_con}) it was assumed
that the strangeness chemical potential $\mu_{s}$ is
zero.

It is convenient to define the effective lepton
density\index{density!lepton!effective} as 
\begin{equation}
 n_{l}\index{$n_{l}$} = n_{le} + n_{l\mu} \qquad [\mbox{fm}^{-3}]
\end{equation}
and the effective lepton chemical 
potential\index{chemical potential!lepton!effective}
\begin{equation}
 \mu_{l}\index{$\mu_{l}$} = \frac{\mu_{le} n_{le} +\mu_{l\mu}
   n_{l\mu}}{n_{l}} \qquad [\mbox{MeV}]
\end{equation}
such that
\begin{equation}
 f(T,n_{b},Y_{q})\index{$f$} =
  -p + \left( \mu_{b} + Y_{q} \mu_{q} \right) n_{b}
\end{equation}
in case of an EoS without leptons and
\begin{equation}
 f(T,n_{b},Y_{q})\index{$f$} =
  -p + \left( \mu_{b} + Y_{l} \mu_{l} \right) n_{b}
\end{equation}
in the case with leptons and charge neutrality\index{charge!neutrality}
($Y_{l} =  Y_{q}$), respectively. 
The tabulated values for 
the entropy density, pressure and chemical potentials
should be given by the first partial derivatives as
\begin{eqnarray}
 s\index{$s$} & = &
 - \left. \frac{\partial f}{\partial T} \right|_{n_{b},Y_{q}} \: ,
 \\
 p\index{$p$} & = &
 n_{b}^{2} \left. \frac{\partial (f/n_{b}) }{\partial n_{b}}
  \right|_{T,Y_{q}} \: ,
 \\
  \mu_{b}\index{$\mu_{b}$} 
  + Y_{i} \mu_{i} 
  & = &
 \left. \frac{\partial f}{\partial n_{b}} \right|_{T,Y_{q}} \: ,
 \\ 
   \mu_{i}  & = & \frac{1}{n_{b}}
 \left. \frac{\partial f}{\partial Y_{i}} \right|_{T,n_{b}} 
\end{eqnarray}
where $i=l(q)$ if leptons are (not) included in the EoS. 
These relations 
will be used to derive the first derivatives of $f$ from
the tabulated thermodynamic quantities $s$, $p$, $\mu_{b}$,
$\mu_{q}$, and $\mu_{l}$ 
in some particular interpolation schemes.

For a mixed second partial derivative, 
the Maxwell relations\index{relation!Maxwell} 
\begin{eqnarray}
 - n_{b}^{2} \left. \frac{\partial (s/n_{b})}{\partial n_{b}}
 \right|_{T,Y_{q}} 
 & = & \left. \frac{\partial p}{\partial T} \right|_{n_{b},Y_{q}} \: ,
 \\
 - \left. \frac{\partial s}{\partial n_{x}} \right|_{T,n_{y},y \neq x}
 & = &  \left. \frac{\partial \mu_{x}}{\partial T} \right|_{n_{y}} \: ,
 \\
  \left. \frac{\partial p}{\partial n_{x}} \right|_{T,n_{y},y \neq x}
 & = &   n_{b} \left. \frac{\partial \mu_{x}}{\partial n_{b}}
 \right|_{T,n_{y}, y \neq b} \: , 
  \\
  \left. \frac{\partial \mu_{x}}{\partial n_{y}} \right|_{T,n_{z},z \neq y}
 & = &  \left. \frac{\partial \mu_{y}}{\partial n_{x}}
 \right|_{T,n_{z}, z \neq x}
\end{eqnarray}
where $x,y,z \in \left\{ b,q,le,l\mu \right\}$ should hold.

The knowlegde of three functions $p=p(T,n_{b},Y_{q})$,
$\mu_{b}\index{$\mu_{b}$}=\mu_{b}(T,n_{b},Y_{q})$, and
$\mu_{q}\index{$\mu_{q}$}=\mu_{q}(T,n_{b},Y_{q})$ or
$\mu_{l}\index{$\mu_{l}$}=\mu_{l}(T,n_{b},Y_{q})$
is sufficient to recover the free energy density
$f\index{$f$}=f(T,n_{b},Y_{q})$
for the particular physical conditions of section \ref{sec:physcon}
($\mu_{s} = 0$ and $n_{q}=n_{l}$).
Further quantities can be derived by
partial derivatives with respect to the parameters
$T$, $n_{b}$ and $Y_{q}$.

Thus it would be convenient
to store only these three quantities in the EoS table\index{table!EoS}
for the thermodynamic properties in order to reduce the memory size.
However, we require that the pressure, the entropy, the three chemical potentials 
and also the free and internal energy densities should be
provided independently in the EoS tables for checking purposes. 
It is also allowed to
store more than these seven basic quantities. 

\section{Structure of tables}
\label{sec:tab_structure}

The CompOSE data base contains for each EoS model data tables
that specify the used \emph{parameter grid} and tables that contain
the \emph{thermodynamic properties, the compositional information}
and \emph{the microscopic information}, respectively.

\subsection{Types of EoS tables}
\label{ssec:tot}

In general, three types of EoS input tables\index{table!input} 
should be available for each EoS
model in order to address different applications:

\begin{enumerate}

\item {\bf Three-dimensional tables}
\index{equation of state!three-dimensional}

These tables depend on all three indepedent parameters $T$, $n_{b}$
and $Y_{q}$. There is only one case:

\begin{itemize}
\item {\bf General purpose EoS table}
\index{equation of state!general purpose}

The recommended parameter\index{parameter!recommended} 
values for the definition of the grid are
given in table \ref{tab:eos_GP}.
This choice of discretation 
corresponds to a table with 
$N_{T}^{\rm max} \times N_{n_{b}}^{\rm max} 
\times N_{Y_{q}}^{\rm max} = 1119720 (1462860) $ data points for
the case a(b) in the $T$ grid.
The table includes neither points with $T=0$~MeV nor
with $Y_{q}=0$.

\end{itemize}

\item {\bf Two-dimensional tables}
\index{equation of state!two-dimensional}

These tables depend on two of the three indepedent parameters $T$, $n_{b}$
and $Y_{q}$. Four cases are considered here:

\begin{itemize}

\item {\bf Zero-temperature EoS table}
\index{equation of state!zero temperature}

The recommended parameter\index{parameter!recommended}
values for the definition of the grid are
given in table \ref{tab:eos_T0}.
This choice of discretization 
corresponds to a table with 
$N_{n_{b}}^{\rm max} \times (N_{Y_{q}}^{\rm max}+1) = 18361$ data points.
Only $T=0$~MeV is considered. Data points for 
$Y_{q}=0$, i.e.\ pure neutron matter,
are included in this EoS table.

\item {\bf Symmetric matter EoS table}
\index{equation of state!symmetric matter}

The recommended parameter\index{parameter!recommended}
values for the definition of the grid are
given in table \ref{tab:eos_NM}.
This choice of discretization 
corresponds to a table with 
$N_{T}^{\rm max} \times N_{n_{b}}^{\rm max} = 18963 (24381) $ data points
for the case a(b) in the $T$ grid.
Only $Y_{q}=0.5$ is considered. $T=0$~MeV, i.e.\ the 
zero-temperature case, is (not) included in this EoS table
for case b(a).

\item {\bf Neutron matter EoS table}
\index{equation of state!neutron matter}

The recommended parameter\index{parameter!recommended}
values for the definition of the grid are
given in table \ref{tab:eos_NM}.
This choice of discretization 
corresponds to a table with 
$N_{T}^{\rm max} \times N_{n_{b}}^{\rm max} = 18963 (24381) $ data points
for case a(b).
Only $Y_{q}=0$ is considered. $T=0$~MeV, i.e.\ the 
zero-temperature case, is (not) included in this EoS table
for case b(a).

\item {\bf EoS table of $\beta$-equilibrated matter}
\index{equation of state!$\beta$ equilibrium}

The recommended parameter\index{parameter!recommended}
values for the definition of the grid are
given in table \ref{tab:eos_NM}.
This choice of discretization 
corresponds to a table with 
$N_{T}^{\rm max} \times N_{n_{b}}^{\rm max} = 18963 (24381) $ data points
for the case a(b) in the $T$ grid.
The charge fraction of strongly interacting particles $Y_{q}$ is determined by the
condition of $\beta$ equilibrium. $T=0$~MeV, i.e.\ the 
zero-temperature case, is (not) included in this EoS table for
case b(a).

\end{itemize}

\item {\bf One-dimensional tables}
\index{equation of state!one-dimensional}

These tables depend only on the parameter $n_{b}$. 
Three cases are considered here:

\begin{itemize}

\item {\bf EoS table of cold symmetric matter}
\index{equation of state!symmetric matter}

The recommended parameter\index{parameter!recommended}
values for the definition of the grid are
given in table \ref{tab:eos_1dim}.
This choice of discretization 
corresponds to a table with 
$N_{n_{b}}^{\rm max} = 301$ data points.
Only $Y_{q}=0.5$ and $T=0$~MeV is considered. 

\item {\bf EoS table of cold neutron matter}
\index{equation of state!neutron matter}

The recommended parameter\index{parameter!recommended}
values for the definition of the grid are
given in table \ref{tab:eos_1dim}.
This choice of discretization 
corresponds to a table with 
$N_{n_{b}}^{\rm max} = 301$ data points.
Only $Y_{q}=0.0$ and $T=0$ is considered. 

\item {\bf EoS table of cold $\beta$-equilibrated matter}
\index{equation of state!$\beta$ equilibrium}

The recommended parameter\index{parameter!recommended}
values for the definition of the grid are
given in table \ref{tab:eos_1dim}.
This choice of discretization 
corresponds to a table with 
$N_{n_{b}}^{\rm max} = 301$ data points.
Only $T=0$~MeV is considered and the charge fraction 
of strongly interacting particles $Y_{q}$
is determined by the conditions of $\beta$-equilibrium and local
charge neutrality.

\end{itemize}

\end{enumerate}

Every set of EoS tables\index{table!EoS} contains up to three tables: the first
that specifies the
thermodynamic\index{table!thermodynamic} 
state of the system (always required), the second that
contains information on the composition\index{table!composition} of the matter
(optional) and the third that gives microscopic 
information\index{table!microscopic} (optional).

The format\index{table!format} of the thermodynamic, the
compositional
and the microscopic
table, respectively, has to follow the following prescriptions.

\subsection{Table with thermodynamic quantities}
\label{ssec:tabthermoquant}
\index{table!thermodynamic}

The first row of this table contains 
three entries: the mass of the neutron $m_{n}$\index{$m_{n}$} 
and that of the proton
$m_{p}$\index{$m_{p}$} in MeV of the particular EoS model and then an integer
$I_{l}$\index{$I_{l}$} that
indicates whether the EoS contains leptons or not. If electrons (and/or
muons) are considered in the EoS, one sets $I_{l}=1$. For other values
of $I_{l}$, it is assumed that there are no electrons (and/or muons)
and the lepton chemical potential $\mu_{l}$ should be set to zero.
The neutron mass $m_{n}$ is used for scaling
certain quantities as specified below.
The following rows in this table contain information
on the particular grid point in the table and the relevant
thermodynamic quantities.
The actually tabulated quantities are given by
\begin{enumerate}
\item {\bf pressure divided by the baryon number 
density $Q_{1}= p/n_{b}$} [MeV], 
\item {\bf entropy per baryon $Q_{2} = s/n_{b}$} [dimensionless],
\item {\bf scaled and shifted baryon chemical potential
    $Q_{3} =\mu_{b}/m_{n}-1$} [dimensionless],
\item {\bf scaled charge chemical potential
    $Q_{4} = \mu_{q}/m_{n}$} [dimensionless],
\item {\bf scaled effective lepton chemical potential
    $Q_{5} = \mu_{l}/m_{n}$} [dimensionless],
\item {\bf scaled free energy per baryon
    $Q_{6} = f/(n_{b}m_{n})-1$} [dimensionless],
\item {\bf scaled internal energy per baryon
    $Q_{7} = e/(n_{b}m_{n})-1$} [dimensionless].
\end{enumerate}
For an ideal gas $Q_{1}$\index{$Q_{i}$} is just the temperature $T$.
The traditional scaling of the chemical potentials and the energies
per baryon
with the neutron mass $m_{n}$\index{$m_{n}$} has been introduced. 

Each line in the thermodynamic table 
(except the first) has to contain the following quantities
\begin{equation}
\label{eq:eos.thermo}
 i_{T}\index{$i_{T}$} \:\: i_{n_{b}}\index{$i_{n_{b}}$} 
 \:\: i_{Y_{q}}\index{$i_{Y_{q}}$} \:\:
 Q_{1} \:\: Q_{2} \:\: Q_{3} \:\: Q_{4} \:\: Q_{5} \:\: Q_{6} \:\:
 Q_{7} \:\:
 N_{\rm add} \:\: 
 \underbrace{q_{1} \:\: 
   q_{2} \:\: \dots
 \:\:  q_{N_{\rm add}}}_{N_{\rm add} \: {\rm quantities}}
\end{equation}
with seven required quantities $Q_{i}$ and 
$N_{\rm add}$\index{$N_{\rm add}$} additional
optional thermodynamic quantities $q_{i}$\index{$q_{i}$}. 
The meaning of the additional quantities\index{quantity!additional}
is specified in the data
sheet\index{data sheet} for each EoS table.
It is assumed that the order of all quantities $q_{i}$ does not change
within an EoS table.
Since the triple of table indices 
$(i_{T},i_{n_{b}},i_{Y_{q}})$
is explicitly specified, the order of tabulation\index{table!order} with
respect to the grid points is not relevant. If the table contains
rows with identical indices $i_{T}$, $i_{n_{b}}$, $i_{Y_{q}}$ then
table entries for the quantities $Q_{i}$, $N_{\rm add}$, 
and $q_{i}$ are used from the row that is read last.

\subsection{Table with the composition of matter}
\label{ssec:compo}
\index{table!composition}

The predicted chemical composition of matter depends strongly
on the employed theoretical model. In principle, the
particle content and hence the data that are required to be
stored can depend on the particular point in the parameter space, too.
Thus the format of the table
has to be adapted to this situation. The essential information is
contained in the particle fractions\index{particle!fraction} 
$Y_{i}$\index{$Y_{i}$} but in some cases
additional information, e.g.\ on the type of phase\index{phase},
might be useful. 

The tabulated quantities in the EoS composition table
are typically given by
\begin{enumerate}
\item {\bf particle fractions} 
  $Y_{i}$\index{$Y_{i}$} [dimensionless],
\item {\bf average mass number of a representative nucleus} 
  $A^{\rm av}$\index{$A^{\rm av}$} [dimensionless],
\item {\bf average charge number of a representative nucleus} 
  $Z^{\rm av}$\index{$Z^{\rm av}$} [dimensionless],
\item {\bf index encoding the type of phase} 
  $I_{\rm phase}$\index{$I_{\rm phase}$} [dimensionless]\\
(The correspondence of the phase index $I_{\rm phase}$ with the 
actual structure of the phase is defined for each EoS model
on the corresponding web page.).
\end{enumerate}


In the following a standard format for the tabulation of compositional
information is defined. It is suitable
for most of the presently available EoS tables of hadronic and quark 
EoS models.
It allows to change 
from point to point the particle fractions
that are given in the table. Thus, it is possible
to select only those particles that are most abundant under the
considered local conditions. Each row of the table  contains the
following entries
\begin{eqnarray}
  \lefteqn{i_{T} \:\: i_{n_{b}} \:\: i_{Y_{q}} \:\: 
 I_{\rm phase} \:\: N_{\rm pairs}
 \underbrace{I_{1} \:\: Y_{I_{1}} \:\: \dots \:\: 
 I_{N_{\rm pairs}} \:\: Y_{I_{N_{\rm pairs}}}}_{N_{\rm pairs} \: {\rm
   pairs}}}
 \\ \nonumber & &
 N_{\rm quad} \:\:
 \underbrace{I_{1} \:\:  A^{\rm av}_{I_{1}}\:\: Z^{\rm
     av}_{I_{1}} \:\: Y_{I_{1}} \:\: \dots \:\: 
   I_{N_{\rm quad}} \:\: A^{\rm av}_{I_{N_{\rm quad}}}\:\: Z^{\rm
     av}_{I_{N_{\rm quad}}} \:\: Y_{I_{N_{\rm quad}}}}_{N_{\rm quad} \: {\rm quadruples}} 
\end{eqnarray}
with $N_{\rm pairs}$\index{$N_{\rm pair}$} 
pairs that combine the particle index $I_{i}$\index{$I_{i}$}
as defined in tables \ref{tab:partindex} and 
\ref{tab:corrindex} and the corresponding
particle fraction $Y_{i}$\index{$Y_{i}$}. 
(Note the definition of particle fractions in equation
\ref{eq:Ydef}.) In addition, there are $N_{\rm quad}$\index{$N_{\rm quad}$}
quadruples that contain an index $I_{i}$  that specifies a group of
nuclei $\mathcal{M}_{I_{i}}$ with average mass number 
\begin{equation}
 A^{\rm av}_{I_{i}}\index{$A^{\rm av}_{I_{i}}$} = \frac{\sum_{j\in \mathcal{M}_{I_{i}}} A_{j}
   Y_{j}}{\sum_{j\in \mathcal{M}_{I_{i}}} Y_{j}} \: , 
\end{equation}
average charge number $Z^{\rm av}_{I_{i}}$\index{$Z^{\rm av}_{I_{i}}$}
\begin{equation}
 Z^{\rm av}_{I_{i}} = \frac{\sum_{j\in \mathcal{M}_{I_{i}}} Z_{j}
   Y_{j}}{\sum_{j\in \mathcal{M}_{I_{i}}} Y_{j}} \: , 
\end{equation}
and combined fraction
\begin{equation}
 Y_{I_{i}}\index{$Y_{I_{i}}$} = \frac{\sum_{j\in \mathcal{M}_{I_{i}}} A_{j}Y_{j}}{A^{\rm
     av}_{I_{i}}}
 = \sum_{j\in \mathcal{M}_{I_{i}}} Y_{j} \: .
\end{equation} 
In case that there are no pairs (quadruples) to be stored, the number
$N_{\rm pairs}$ ($N_{\rm quad}$) has to be set to zero.
The average mass and charge numbers correspond to those
of a representative heavy nucleus if there is only one group of nuclei
except the lightest that are considered explicitly. 
However, it is also possible to define several subsets of nuclei
with corresponding average mass numbers, charge numbers and fractions.
The correlation between the index $I_{i}$ and the set of
nuclei $\mathcal{M}_{I_{i}}$
is defined for each EoS model separately and given in the
data sheet\index{data sheet} on the CompOSE web pages.
Since the composition of matter can vary rapidly with a change of the
parameters, it is not required that the fractions of all particles or
particle sets have to given in reach row. It is sufficient to list
only those that are relevant, e.g.\ with positive $Y_{i}$. Others can
be omitted.

\subsection{Table with information on microscopic quantities}
\label{ssec:micro}
\index{table!microscopic}

Besides the information on thermodynamic and compositional
quantities, most EoS models can provide additional information
on microscopic quantities\index{quantity!microscopic} 
of an individual particle $i$, e.g.\
\begin{enumerate}
\item {\bf Landau effective mass divided by the particle mass} 
  $m^{L}_{i}/m_{i}$\index{$m^{L}_{i}$} [dimensionless],
\item {\bf Dirac effective mass divided by the particle mass} 
  $m^{D}_{i}/m_{i}$\index{$m^{D}_{i}$} [dimensionless],
\item {\bf non-relativistic single-particle potential} 
  $U_{i}$\index{$U_{i}$} [MeV],
\item {\bf relativistic vector self-energy} 
  $V_{i}$\index{$V_{i}$} [MeV],
\item {\bf relativistic scalar self-energy} 
  $S_{i}$\index{$S_{i}$} [MeV],
\item {\bf pairing gap}
  $\Delta_{i}$\index{$\Delta_{i}$} [MeV].
\end{enumerate}
These quantities are stored in an additional table 
if available.
In this case, each line has the format
\begin{equation}
 i_{T}\index{$i_{T}$} \:\: i_{n_{b}}\index{$i_{n_{b}}$} 
 \:\: i_{Y_{q}}\index{$i_{Y_{q}}$} \:\:
 N_{\rm qty} \:\: 
 \underbrace{K_{1} \:\: q_{K_{1}} \:\: K_{2} \:\:
   q_{K_{2}} \:\: \dots
 \:\: K_{N_{\rm qty}} \:\: q_{K_{N_{\rm qty}}}}_{N_{\rm qty} \: {\rm
   pairs}}
\end{equation}
where $N_{\rm qty}$\index{$N_{\rm qty}$} 
is the number of stored quantities $q_{K_{i}}$.
The composite index\index{index!composite} 
$K_{i}$\index{$K_{i}$} identifies uniquely both the particle or
correlation and the physical quantity. It is formed as
\begin{equation}
 K_{i} = 1000 \: I_{i} + J_{i}
\end{equation}
with the particle or correlation
index\index{index!particle}\index{index!correlation} 
$I_{i}$\index{$I_{i}$} from table
\ref{tab:partindex} or \ref{tab:corrindex}, respectively, and
the quantity index $J_{i}$\index{$J_{i}$} from table \ref{tab:ident_micro}.

\subsection{Tables with parameters}
\label{ssec:paratables}
\index{table!parameter}

The mesh points for the general purpose EoS 
are stored in three individual data files that are
used as an input in addition to the three tables containing
the thermodynamic, compositional and microphysical information.
It is assumed that the discretization scheme 
for the relevant parameters in the zero-temperature
and in the neutron-matter table is identical 
to the grid for the general purpose table.
Parameter values should be given with at least eight significant
digits. For the parameter temperature, the corresponding
table contains a single
entry in each row. In the first row, 
$N_{T}^{\rm min}$\index{$N_{T}^{\rm min}$} is given,
and $N_{T}^{\rm max}$\index{$N_{T}^{\rm max}$} 
in the second row. Then the parameter values
$T(N_{T}^{\rm min})$, $T(N_{T}^{\rm min}+1)$, \dots, $T(N_{T}^{\rm max})$, 
are given in every following row. In total, the table contains
$N_{T}^{\rm max}-N_{T}^{\rm min}+3$ rows. The same storage scheme
applies for the parameters $n_{b}$ and $Y_{q}$.

\subsection{Identification of tables}
\label{ssec:identtables}
\index{table!identification}

Each Eos table is identified by a unique name and a unique extension
that correspond to the model and type of table, respectively. 
Using the generic name {\tt eos} for a particular EoS model, there are 
always three parameter tables
\begin{itemize}
 \item {\tt eos.t}\index{eos.t} (required)
 \item {\tt eos.nb}\index{eos.nb} (required)
 \item {\tt eos.yq}\index{eos.yq} (required)
\end{itemize}
that define the discretization mesh for the three parameters
$T$, $n_{b}$ and $Y_{q}$ as discussed in the previous section.
All three tables are required even for EoS tables with less than three
dimensions.

The thermodynamic, compositional and microscopic quantities are
stored in three tables with the generic names
\begin{itemize}
 \item {\tt eos.thermo}\index{eos.thermo} (required)
 \item {\tt eos.compo}\index{eos.compo} (optional)
 \item {\tt eos.micro}\index{eos.micro} (optional)
\end{itemize}
where only the first one is required for every EoS model since
information
on the composition or microscopic details are not always available. 

\subsection{Precision of tabulated quantities}

EoS quantities of a particular model have a limited
precision\index{precision} in practical calculations. Maschine precision is hardly ever
reached. The format\index{table!format} 
of the tabulated data should reflect this fact and
only the significant number of digits should be given to avoid excessive
memory size. Usually eight significant digits are sufficient since
the error in the interpolation can be considerably larger.

\section{Data sheet}

Information on an EoS that is not provided directly in the tables but relevant
for the characterisation of the particular model can be found in a
data sheet\index{data sheet} that comes along with the tables on the
CompOSE web pages\index{web page}. Some quantities are extracted
automatically from the tabulated data, others need to be specified by
the contributor. E.g., characteristic nuclear matter parameters (see
section \ref{sec:nucmatpar}) are obtained by using the program {\tt
compose.f90} (see section \ref{sec:data}) if an EoS table for pure
nuclear matter without electrons or muons is available. 
Similarly, information on the parameter grid, type and dimension of
the Eos table, considered particles and additionally stored quantities
are extracted by the program. The EoS of
stellar matter in 
$\beta$-equilibrium\index{equation of state!$\beta$ equilibrium} 
is generated by using
{\tt compose.f90}\index{compose.f90} if the EoS table for matter with electrons of muons
is provided by the contributor. From this extracted EoS table,
characteristic properties of neutron stars, e.g.\ 
maximum masses\index{maximum mass}, are 
calculated and given in the data sheet. It also contains information
on the considered constituent particles in the model. The contributor
has to supply a short characterisation of the EoS model and relevant
references\index{references}. He/she also needs to define the meaning of the index
$I_{\rm phase}$\index{$I_{\rm phase}$} for the phases\index{phase} considered in the EoS.


\newpage

\begin{table}[t]
\begin{center}
\caption{\label{tab:eos_GP}%
Parameter values for the recommended\index{parameter!recommended} 
discretization in
$T$, $n_{b}$ and $Y_{q}$ for a general purpose EoS table.}
\begin{tabular}{lcccc}
 \toprule
 quantity & reference & minimum & maximum  & resolution \\
                      & value(s) & index & index & parameter \\
 \midrule
 $T$, case a & $T^{\rm ref} = 0.1$~MeV & 
 $N_{T}^{\rm min} = 1$ & $N_{T}^{\rm max} = 81$ & $M_{T}
 = 25$ \\
 $T$, case b & $T^{\rm ref} = 0.1$~MeV & 
 $N_{T}^{\rm min} = 1$ & $N_{T}^{\rm max} = 62$ & $M_{T}=25$ \\
 $n_{b}$ & $n_{b}^{\rm ref} = 10^{-12}$~fm${}^{-3}$ & 
 $N_{n_{b}}^{\rm min} = 1$ & $N_{n_{b}}^{\rm max} = 301$ & $M_{n_{b}}
 = 25$ \\
 $Y_{q}$ & N/A &
 $N_{Y_{q}}^{\rm min} = 1$ & $N_{Y_{q}}^{\rm max} = 60$ & $M_{Y_{q}} = 100$ \\
 \bottomrule
\end{tabular}
\end{center}
\end{table}

\begin{table}[t]
\begin{center}
\caption{\label{tab:eos_T0}%
Parameter values for the recommended\index{parameter!recommended} 
discretization in
$n_{b}$ and $Y_{q}$ for a zero-temperature EoS table.}
\begin{tabular}{lcccc}
 \toprule
 quantity & reference & minimum & maximum  & resolution \\
                      & value & index & index & parameter \\
 \midrule
 $n_{b}$ & $n_{b}^{\rm ref} = 10^{-12}$~fm${}^{-3}$ & 
 $N_{n_{b}}^{\rm min} = 1$ & $N_{n_{b}}^{\rm max} = 301$ & $M_{n_{b}}
 = 25$ \\
 $Y_{q}$ & N/A &
 $N_{Y_{q}}^{\rm min} = 0$ & $N_{Y_{q}}^{\rm max} = 60$ & $M_{Y_{q}} = 100$ \\
 \bottomrule
\end{tabular}
\end{center}
\end{table}

\begin{table}[t]
\begin{center}
\caption{\label{tab:eos_NM}%
Parameter values for the recommended\index{parameter!recommended} 
discretization in
$T$ and $n_{b}$ for EoS tables of symmetric matter, neutron matter or
$\beta$-equilibrated matter.}
\begin{tabular}{lcccc}
 \toprule
 quantity & reference & minimum & maximum  & resolution \\
                      & value & index & index & parameter \\
 \midrule
 $T$, case a & $T^{\rm ref} = 0.1$~MeV & 
 $N_{T}^{\rm min} = 1$ & $N_{T}^{\rm max} = 81$ & $M_{T}
 = 25$ \\
 $T$, case b & $T^{\rm ref} = 0.1$~MeV & 
 $N_{T}^{\rm min} = 0$ & $N_{T}^{\rm max} = 62$ & $M_{T} = 25$ \\
 $n_{b}$ & $n_{b}^{\rm ref} = 10^{-12}$~fm${}^{-3}$ & 
 $N_{n_{b}}^{\rm min} = 1$ & $N_{n_{b}}^{\rm max} = 301$ & $M_{n_{b}}
 = 25$ \\
 \bottomrule
\end{tabular}
\end{center}
\end{table}

\begin{table}[t]
\begin{center}
\caption{\label{tab:eos_1dim}%
Parameter values for the recommended\index{parameter!recommended} 
discretization in
$n_{b}$ for one-dimensional EoS tables.}
\begin{tabular}{lcccc}
 \toprule
 quantity & reference & minimum & maximum  & resolution \\
                      & value & index & index & parameter \\
 \midrule
 $n_{b}$ & $n_{b}^{\rm ref} = 10^{-12}$~fm${}^{-3}$ & 
 $N_{n_{b}}^{\rm min} = 1$ & $N_{n_{b}}^{\rm max} = 301$ & $M_{n_{b}}
 = 25$ \\
 \bottomrule
\end{tabular}
\end{center}
\end{table}

\chapter{Extending CompOSE}
\label{ch:extensions}

CompOSE is meant to be under constant development.
The main aim is to enlarge the data base by adding
EoS tables of more and more models. 
The most simple way is to convert tables of existing model
calculations to the generic CompOSE scheme of tabulation in
the data files.
A more involved task is to develop a new EoS model and to provide the
results in the appropriate format. In either case, we ask
you to contact the CompOSE core team by sending an email to
\begin{quote}
{\tt develop.compose@obspm.fr}. 
\end{quote}
You will be contacted and all
questions will be clarified 
on how your results can be incorporated in the CompOSE data base
and made accessible to the public.

For the future it is foreseen that the functionality of the CompOSE
database will be extended in order to make even more results
of EoS model calculations available that are not included in
the present sets of data. It is also envisioned to add more quantities
to the data tables that might be relevant for astrophysical
applications and beyond.







\part{For Users}
\label{part:user}

\chapter{Models for the equation of state}
\label{ch:eosmodels}
There is a large number of model equations of state. They fall into
different classes 
depending on the theoretical approach.
We use a rough classification\index{equation of state!classifiation} 
scheme that takes the type
of model and complexity into account.
The main difficulty in generating an equation of state is the
description of the strongly interacting constituents, hadrons and/or
quarks. 

Details and characteristic parameters of each particular EoS
can be found in a data sheet that accompanies the data tables
on the CompOSE web site 
\begin{quote}
 \url{http://compose.obspm.fr}.
\end{quote}

For electrons\index{electron} and muons\index{muon} 
a uniform Fermi gas\index{Fermi gas} model is
employed in many cases. This assumption simplifies the calculation significantly
in case of spatially inhomogeneous charge distributions.
Details on the treatment of electrons and muons 
are given for each EoS separately on the data sheet.

At positive temperatures photons\index{photon} contribute to the thermodynamic
properties of the system. Their treatment is discussed in section
\ref{sec:photons}. For each EoS in the database a remark is given
whether photons are considered or not.
 
Before theoretical models for the EoS
are summarized, the definition of commonly used
nuclear matter parameters\index{parameter!nuclear matter} 
is given in section \ref{sec:nucmatpar}.
They characterize the essential
properties of nuclear matter and can serve
in a first comparison of the various models.
Be, however,
careful since these parameters are only valid in the vicinity of
saturation for symmetric, i.e.\ same number of protons and neutrons, matter.
Specific values of these parameters are given in the data sheet
for the individual models.

In section \ref{sec:eosmodels} a general overview on the most
common approaches to describe the EoS of nuclear matter is
presented and their major features are summarized. 

Some comments on phase transitions within the EoS\index{phase!transition} 
are presented in section \ref{sec:phasetransitions}.
In the last section 
a few remarks of caution are given.

\section{Nuclear matter properties}
\label{sec:nucmatpar}
Realistic models of nuclear, i.e.\ purely hadronic, matter\index{matter!hadronic} 
without leptons show
some general features that are related to the occurence of the
saturation\index{saturation} 
phenomenon. Due to the symmetries of the strong interaction
uniform nuclear matter\index{matter!uniform} 
at zero temperature reaches a state with the
largest binding energy per nucleon
at a finite saturation density\index{density!saturation} 
$n_{b}^{\rm sat}$\index{$n_{b}^{\rm sat}$} with equal concentrations
of neutrons and protons\footnote{This is exactly true only if the mass
  difference between neutrons and protons is neglected.}. 
For $T=0$~MeV, the energy per nucleon\index{energy!per nucleon} 
$\mathcal{E}$\index{$\mathcal{E}$}
can be considerd as a function of the baryon number density $n_{b}$
and the asymmetry\index{asymmetry}
\begin{equation}
 \alpha\index{$\alpha$} = \frac{n_{n}-n_{p}}{n_{n}+n_{p}} = 1-2Y_{q} \qquad
 [\mbox{dimensionless}]
\end{equation}
that vanishes for symmetric nuclear matter, i.e.\  $\alpha = 0$ and
$Y_{q} = 0.5$. It can be
expanded in a power series around the saturation point as
\begin{eqnarray}
\label{eq:Eseries}
 \mathcal{E}(n_{b},\alpha) & = &
 -B_{\rm sat} + \frac{1}{2} K x^{2} + \frac{1}{6}
 K^{\prime} x^{3} + \dots 
 \\ \nonumber & & 
 + \alpha^{2} \left( J +  L x + \frac{1}{2} K_{\rm sym}
   x^{2} + \dots \right) + \dots 
 \qquad [\mbox{MeV}]
\end{eqnarray}
with the relative deviation of the actual density from the saturation
density
\begin{equation}
\label{eq:def_x}
 x\index{$x$} = \frac{1}{3} \left( 
 \frac{n_{b}}{n_{b}^{\rm sat}} -1  \right) \qquad [\mbox{dimensionless}]
\end{equation}
and the asymmetry $\alpha$. The coefficients 
$B_{\rm sat}$\index{$B_{\rm sat}$},
$K$\index{$K$}, $K^{\prime}$\index{$K^{\prime}$}, 
$J$\index{$J$}, $L$\index{$L$}, $K_{\rm sym}$\index{$K_{\rm sym}$} [MeV], 
\dots and $n_{b}^{\rm  sat}$\index{$n_{b}^{\rm sat}$} [fm${}^{-3}$]
characterize the behaviour of the EoS.
They can, of course, only give an indication of the general features
of the EoS, since these
quantities are relevant essentially in the vicinity of the saturation
density\index{density!saturation} at zero
temperature and close to symmetric nuclear matter (equal number
of protons and neutrons).  Extrapolations based on this polynomial
expansion might be dangerous. 
The numerical factor in
equation (\ref{eq:def_x}) 
is of historical origin since 
the Fermi momentum\index{momentum!Fermi} 
$k_{F}$\index{$k_{F}$} [fm$^{-1}$] instead of the density
$n_{b} = k_{F}^{3}/\pi^{2}$ of symmetric nuclear matter at zero
temperature was originally
used as the expansion parameter. 
Note that contributions linear in
$\alpha$ and linear in $x$ for $\alpha=0$ are absent in the expansion
due to the condition of minimum energy per baryon.

In order to study the dependence on the asymmetry $\alpha$, it is also
convenient to introduce the symmetry energy\index{energy!symmetry}
\begin{equation}
 E_{\rm sym}\index{$E_{\rm sym}$}(n_{b}) = \left.
 \frac{1}{2} \frac{\partial^{2} \mathcal{E}(n_{b},\alpha)}{\partial
   \alpha^{2}} \right|_{\alpha = 0} \qquad [\mbox{MeV}]
\end{equation}
that is a function of the baryon number density $n_{b}$ only.
In many cases the quadratic approximation\index{approximation!quadratic} 
in $\alpha$ in equation 
(\ref{eq:Eseries}) is sufficient. Then the symmetry energy can also
be calculated from the finite difference approximation
\begin{equation}
 E_{\rm sym}(n_{b}) = \frac{1}{2}
 \left[ \mathcal{E}(n_{b},-1)  - 2 \mathcal{E}(n_{b},0) 
 + \mathcal{E}(n_{b},1)\right]  \qquad [\mbox{MeV}]
\end{equation}
comparing symmetric matter with pure neutron and pure proton matter.

In the following, the coefficients appearing in equation (\ref{eq:Eseries})
are discussed in more detail and typical values are given.
\begin{itemize}
\item The {\bf saturation density}\index{density!saturation} 
$n_{b}^{\rm sat}$\index{$n_{b}^{\rm sat}$} [fm${}^{-3}$]
  of symmetric nuclear
  matter is defined by the condition that the pressure\index{pressure} 
  vanishes, i.e.\
  \begin{equation}
    p\index{$p$} = \left. n_{b}^{2} \frac{d\mathcal{E}(n_{b},0)}{dn_{b}}
    \right|_{n_{b}=n_{b}^{\rm sat}} = 0 \qquad [\mbox{MeV~fm}^{-3}]
  \end{equation}
  and the energy per baryon becomes minimal.
  Typical values are in the range 
  $0.15~\mbox{fm}^{-3} < n_{b}^{\rm sat} < 0.17~\mbox{fm}^{-3}$ \cite{DanLee}.
\item The {\bf binding energy at saturation}\index{energy!binding} 
  $B_{\rm sat}$\index{$B_{\rm sat}$} [MeV]
  of symmetric nuclear matter typically
  lies in the range 
  $15.6~\mbox{MeV} < B_{\rm sat} < 16.2~\mbox{MeV}$ \cite{DanLee}.
  This quantity can be obtained from Bethe-Weizs\"{a}cker mass
  formulas\index{mass formula!Bethe-Weizs\"{a}cker} 
  for nuclei ${}^{A}{Z}$ by an extrapolation of $A=2Z$
  to infinity.
\item The {\bf incompressibility of bulk nuclear matter}\index{incompressibility}
  $K$\index{$K$} [MeV] quantifies the curvature\index{curvature} 
  of the binding energy per baryon
  with respect to the density variation at saturation
  since it is defined by
  \begin{equation}
    K\index{$K$} =  \left. 9 n_{b}^{2} \frac{\partial^{2} 
        \mathcal{E}(n_{b},0)}{\partial n_{b}^{2}} 
    \right|_{n_{b}=n_{b}^{\rm sat}} 
    =  \left. 9 n_{b} \frac{\partial (p/n_b)}{\partial n_{b}}
\right|_{n_{b}=n_{b}^{\rm sat}} 
\qquad [\mbox{MeV}] \: .
  \end{equation}
  It is related to the isothermal
  compressibility\index{compressibility!isothermal} 
  $\kappa_{T}$\index{$\kappa_{T}$} (\ref{eq:kappat}) at zero temperature, 
  zero asymmetry and saturation density by
  \begin{equation}
   K = \frac{9}{\kappa_{T} n_{b}^{\rm sat}} \qquad [\mbox{MeV}] \: .
  \end{equation}
  Thus $K$ would be better called compression
  modulus\index{compression modulus}.
  Today, the preferred value for $K$ from nuclear experiments is
  $240 \pm 10 $~MeV \cite{Piekarewicz:2009gb}. It can be derived,
  e.g.\ from studies of 
  isoscalar giant monopole   excitations of nuclei. 
  Obviously, the given error is
  rather small, such that many EoS should be disfavored because the value of
  $K$ lies well outside the range given above. We will, however, keep these
  EoS in our data base for two reasons. The first one is purely historical: in
  many simulations EoS with values for the incompressibility outside the range
  given above have been used, such that for comparison with the existing
  literature it is interesting to have  tables at hand with these values. The
  most prominent example is perhaps the Lattimer-Swesty
  EoS~\cite{Lattimer:1991nc} with the parameter set where $K = 180$~MeV.  The
  second one is that the experimental result of $240\pm 10$~MeV is not
  uncontested. In particular, the extraction of this value from data on
  isoscalar giant monopole 
  resonances\index{giant resonance!isoscalar monopole} 
  depends on the density dependence of the
  nuclear symmetry energy\index{energy!symmetry}, 
  a quantity under intensive debate in recent
  years. We therefore think that a wider range of values should be considered
  and it can give an indication about the uncertainties in the simulations
  coming from the uncertainties in the EoS.
\item The {\bf skewness coefficient of bulk nuclear matter}\index{skewness}
  $K^{\prime}$ [MeV] is defined by the third derivative of the energy
  per baryon as
  \begin{equation}
    K^{\prime}\index{$K^{\prime}$} = 27 n_{b}^{3} \left. \frac{\partial^{3} 
        \mathcal{E}(n_{b},0)}{\partial n_{b}^{3}} \right|_{n_{b} =
      n_{b}^{\rm sat}} 
 = 27 n_{b}^{2} \left. \frac{\partial^{2} 
        (p/n_{b})}{\partial n_{b}^{2}} \right|_{n_{b} = n_{b}^{\rm
        sat}}  - 6 K
 \qquad [\mbox{MeV}] \: .
  \end{equation}
  The value of $K^{\prime}$ in combination with $K$ determines the surface
  properties of nuclei, e.g.\ the ratio of the surface tension to the
  surface thickness. Thus, in order to fix $K^{\prime}$ in a particular model,
  it is important to include quantities like radii and the surface thickness
  in the procedure to fit the model parameters.
\item The {\bf symmetry energy at saturation}\index{energy!symmetry} $J$ [MeV]
  is just given by
 \begin{equation}
   J\index{$J$} = E_{\rm sym}(n_{b}^{\rm sat}) \qquad [\mbox{MeV}].
 \end{equation}
  This coefficient mainly determines the isospin\index{isospin} 
  dependence of the 
  binding energy of nuclei. It is important in predicting
  masses of exotic nuclei far away from the valley of 
  stability\index{valley of stability}
  in the chart of nuclei.
  Recently there has been a lot of
  discussion about its value, the commonly assumed range is 
  26 MeV $\lsim J \lsim $ 36
  MeV~\cite{DanLee,Chen:2010qx,Carbone:2010az,Klimkiewicz:2007zz,Xu:2010fh,Lat12,Tsa12}. 
\item The {\bf symmetry energy slope
    coefficient}\index{energy!symmetry!slope} 
  $L$ [MeV] that
  is obtained from
  \begin{equation}
    L\index{$L$} = 3 n_{b} \left. \frac{dE_{\rm sym}(n_{b})}{dn_{b}} \right|_{n_{b} =
      n_{b}^{\rm sat}} \qquad [\mbox{MeV}].
  \end{equation}
  This coefficient is related to the density dependence of the neutron
  matter EoS near the saturation density. 
  It is strongly correlated with the neutron skin
  thickness\index{skin thickness!neutron} 
  of heavy nuclei like ${}^{208}$Pb. Unfortunately,
  experimentally determined values of the neutron skin thickness are not very
  precise so far and a large range of values for $L$ is found by comparing
  different EoS models.
  Typical values lie in the range
28 MeV $\lsim L \lsim$ 100 MeV~\cite{Chen:2010qx,Carbone:2010az,%
Klimkiewicz:2007zz,Xu:2010fh,Lat12,Tsa12,Centelles:2008vu}.
\item The {\bf symmetry incompressibility}\index{incompressibility!symmetry}
  $K_{\rm sym}$ [MeV] 
  quantifies the curvature of the symmetry energy
  with respect to the density variation at saturation.
  It is defined by
  \begin{equation}
    K_{\rm sym}\index{$K_{\rm sym}$} = 9 n_{b}^{2} \left. \frac{d^{2} 
        E_{\rm sym}}{d n_{b}^{2}} \right|_{n_{b} =
      n_{b}^{\rm sat}} \qquad [\mbox{MeV}] \: .
  \end{equation}
  Typical values lie in the range $-500~\mbox{MeV} 
  < K_{\rm sym} < 100~\mbox{MeV}$ \cite{DanLee}.
\end{itemize}
For each EoS in the CompOSE data base, the actual values of the nuclear
matter coefficients are given if available.

\section{Overview of models}
\label{sec:eosmodels}
Models for the equation of state of dense matter can be grouped into
three major categories concerning the fundamental constituents. In
each category there are models that are more or less microscopic
using different approximations. In the following a short overview
is given. Details on a particular EoS model can be found on the 
corresponding web pages of the CompOSE web site.
Of course, there are cases where it is not possible to sort a
particular model uniquely into a category because concepts of
different models are combined.

\begin{enumerate}
\item{Hadronic models}\index{model!hadronic}
 \begin{enumerate}
 \item{Statistical models:}\index{model!statistical}
 e.g.\ nuclear statistical equilibrium (NSE)\index{NSE} models,
 virial equation of state (VEoS)\index{VEoS}
 \item{Density functional models:}\index{model!density functional}
 e.g.\ phenomenological parameter functional models,
 Skyrme-Hartree-Fock (SHF)\index{SHF} models,
 relativistic mean-field (RMF)\index{RMF} models
 \item{Models with realistic microscopic potentials:}
 e.g.\ (Dirac) Brueckner-Hartree-Fock [(D)BHF]\index{BHF}\index{DBHF} models,
 antisymmetrized or fermionic molecular dynamics (AMD, FMD)\index{AMD}\index{FMD} models, 
 quantum Monte Carlo (QMC)\index{QMD} models
 \end{enumerate}
\item{Quark models}\index{model!quark}
 \begin{enumerate}
  \item{Bag models:}\index{model!bag}
 e.g.\ MIT bag models,
 thermodynamic bag models
  \item{(Polyakov loop) Nambu--Jona-Lasinio [(P)NJL]\index{NJL}\index{PNJL} models}
 \end{enumerate}
\item{Hybrid models}\index{model!hybrid}
\end{enumerate}

\section{Phase transitions}
\label{sec:phasetransitions}

Many EoS models exhibit features that require the construction of a
phase transition\index{phase!transition} with a coexistence region of different phases in
certain regions of the parameter space. This is the consequence of
thermodynamic rules that require the system to take on a particular
state in equilibrium respecting the convexity/concavity of a certain
thermodynamic potential such that a thermodynamically
stable minimim/maximum is reached.  This depends on the supposed conditions
and chosen parameters.  A prominent example is the 
van-der-Waals\index{equation of state!van-der-Waals}
equation of state with the occurence of a 
liquid-gas phase transition\index{phase!transition!liquid-gas}
below a critical temperature that connects a low-density gas phase
with a high-density liquid phase. Pure nuclear matter without leptons
and in the absence of electromagnetic interactions exhibits such a
liquid-gas type first order phase
transition\index{phase!transition!first order}. In the case of dense 
stellar matter the situation is more complicated due to several
facts. First of all, electrons and the electromagnetic interaction cannot be
neglected. This means that at the transition from inhomogeneous
matter at low densities containing nuclei to a homogeneous phase with
nucleons at high densities, there is a formation of spatial
inhomogeneities at microscopic scales and there is no first order phase
transition. The thermodynamic quantities are perfectly continuous
(see e.g.~\cite{Raduta10}). It can of course
be appealing for simplicity, to describe this transition 
within a model still by a first order phase transition, but it has to be
kept in mind that this is an approximation. The second point is that
the relevant degrees of freedom can change within the parameter
space, e.g.\ from nuclei to nucleons and even to quarks.

In general, the correct approach to a phase transition in an
equilibrated system is given by the so-called Gibbs
construction\index{Gibbs construction}.
It requires the equality of all intensive variables of the relevant
thermodynamic potential on the two points of the coexistence manifold in the
parameter space that connect the two phases.
Technically, this can be achieved in various ways and the final
EoS should be independent of the construction. However, in many EoS tables,
only simplified phase transition constructions are used, if considered
at all. In the present version of CompOSE, all EoS tables are included
as given by the authors without any discrimination with respect to the
applied construction of the phase transition. 
In a future version of CompOSE it is planned
to offer the computational tools to construct phase transitions on top
of the available EoS tables.

\section{Remarks of caution}
\label{sec:remarks}
We provide equations of state tables
in the CompOSE data base that cover large ranges
in temperature, baryon number density and charge fraction of strongly
interacting particles.
However, we cannot guarantee that the physical description of 
matter under these conditions is close to reality. The user
has to judge whether a particular EoS is suitable for her or his
application.

The user also has to keep in mind that the thermodynamic and
compositional properties of nuclear matter\index{matter!nuclear} without electrons (and
muons) and dense stellar matter\index{matter!stellar} 
with electrons (and muons) are rather
different because in the former case the Coulomb interaction is
artificially neglected. This can have drastic consequences for
the occurence and type of phase transitions. Thus it is not
recommended to generate a dense matter EoS from a nuclear matter EoS
simply by adding the contribution from electrons (and muons).

In many cases it is desirable to extend a particular EoS to densities
and temperatures 
below or above the available or recommended
parameter ranges\index{parameter!range}. 
In these regimes, an
equation of state depending only on temperature, charge
fraction of strongly interacting particles and baryonic density to describe matter 
might not be sufficient, e.g.\ at very low densities and temperatures
where the time scales of 
thermodynamic equilibration\index{equilibrium!thermodynamic} 
reaction rates become large
and full reaction networks have to be considered. 
However, for many purposes a detailed description of matter
in these regime is not necessary, such that we decided to furnish
tables for these conditions, too. 

Our knowledge about the QCD\index{QCD} 
phase diagram\index{phase!diagram} suggests that there could be
a transition from a
hadronic phase\index{phase!hadronic} to a
quark-gluon plasma\index{quark-gluon plasma} 
within the range of densities and temperatures
reachable in core-collapse supernovae\index{supernova!core-collapse}, 
hence within the range of our
tables. Of course, there are lots of uncertainties about this phase
transition\index{phase!transition}, 
so that its occurence cannot be affirmed, but the
possibility has to be kept in mind when employing a purely hadronic
EoS\index{equation of state!hadronic}
up to densities well above nuclear matter 
saturation density\index{density!saturation} and
temperatures as high as several tens of MeV. Within CompOSE there will be
tables available including this phase transition.
Even without thinking about a
QCD phase transition, other forms of (exotic) matter shall appear at
high densities and temperatures. Already for cold matter\index{matter!cold} EoS
used for neutron star\index{star!neutron} 
modelisation for a long time, hyperons\index{hyperon}, pions\index{pion} and
kaons\index{kaon} 
have been considered. At temperatures above about 20~MeV, this point
becomes even more crucial. Thus using a purely nuclear EoS in this
regime can be a first approximation, technically very appealing, but
again it has to be kept in mind that probably the 
considered physics is too poorly described.

\chapter{Online services and data handling}
\label{ch:online-service}
\section{User registration}
If you want to use the EoS data and routines that are provided
by CompOSE through the online service\index{online service}, 
you should register to our system. 
Registration\index{registration} is simple and 
helps us to see who uses our services. It also avoids traffic.
The tables and codes are offered free of charge but come without warranty.
Please acknowledge and give proper reference to CompOSE if you use our service
for your applications, presentations and publications. 

For the purpose of registration use the registration form 
that can be accessed through
the CompOSE homepage\index{homepage} at 
\begin{quote}
{\tt http://compose.obspm.fr}.
\end{quote}
After approval by one of the administrators\index{administrator} 
maintaining the site you will 
receive your personal access data via email and are ready to use the 
CompOSE data base for generating your individual EoS tables.


\section{EoS data sheets}

Each EoS of the CompOSE data base is accompanied with a data
sheet\index{data sheet}
available for download from the web site. The data sheet provides
essential information that allows the user to decide whether the EoS 
is suitable for her/his application. It contains information on
the origin and creation of the EoS table, a short abstract of the
physical model, references, the parameter ranges and considered
particle species, a summary of the available data files and fundamental quantities
that characterize the EoS such as nuclear matter and
neutron star properties, if available.

\section{Options for using EoS data}
\label{sec:eos}
There are different options for
downloading and/or generating EoS data and 
tables\index{equation of state!table}. 
It is possible to obtain EoS tables
for different models and for different purposes.   
The original EoS tables are given as plain ASCII files that allow
every user to read the data without the need for further codes.
The general idea is that
a user chooses a particular model from those available on
the CompOSE web site\index{web site}.
Then, there are two different ways of accessing EoS data:
\begin{enumerate}
\item The user downloads data tables of an available EoS in the original
form and uses her or his own routines to handle the data.
For every model EoS 
there are three tables that contain the details of the discretization
of the parameters: {\tt eos.t}\index{eos.t}, {\tt  eos.nb}\index{eos.nb} 
and {\tt eos.yq}\index{eos.yq}.
The original EoS data are stored as functions of 
temperature\index{temperature} $T$\index{$T$},
baryon number density\index{density!number!baryon}
$n_{b}$\index{$n_{b}$} 
and charge fraction of strongly interacting particles\index{fraction!charge} 
$Y_{q}$\index{$Y_{q}$}.
The last quantity is identical to the electron
fraction\index{fraction!electron} 
$Y_{e}$\index{$Y_{e}$} if electrons are
the only charged leptons considered in the EoS model.
The abbreviation {\tt eos} is the name of a particular EoS identifying
the model and the type of table uniquely. 
The actual data on the thermodynamic,
compositional and microscopic properties are stored in the files
{\tt eos.thermo}\index{eos.thermo}, 
{\tt eos.compo}\index{eos.compo} and 
{\tt eos.micro}\index{eos.micro}, respectively,
as far as available for that particular model. See section \ref{sec:tab_structure}
for the format of all the tables. 
\item The user downloads the original data tables for a particular
EoS as described above and, in addition, 
several files 
that contain routines for reading, testing, interpolating and transforming
the data. These codes serve four major purposes:
\begin{itemize}
\item to interpolate the original EoS data tables in order to obtain
the quantities at parameter values different from the original tabulation,
\item to calculate additional quantities that are not given in the
original data files,
\item to select those quantities that are relevant for a particular
application and to store them in separate data files in a format
more convenient for the user,
\item to provide EoS data tables in the advanced 
 HDF5 format\index{format!HDF5} (see\\
\url{http://www.hdfgroup.org/HDF5/}) that is widely used in the
astrophysics community.
\end{itemize}
Details on how to work with the subroutines are described in 
the next section. 
\end{enumerate}

\section{Handling and transforming EoS data}
\label{sec:data}

The handling of EoS data is considerably simplified by using the 
program files as such or the
subroutines included in these files. There are only few steps required
in order to generate
customized data tables from the original sets of EoS tables.
In addition to these data files, the user has to provide 
input files\index{file!input} that contain the control parameters if the
program {\tt compose}\index{compose}, generated from the
program files, is used directly. Alternatively,
the parameters can be given as arguments of the various subroutines.

All quantities describing the thermodynamic, compositional and
microscopic properties of the matter are found for arbitrary values of the
table parameters with an interpolation\index{interpolation} scheme
that is described in detail in appendix~\ref{sec:interpol}.
It is possible to create tables with different mesh settings and ranges.
Of course, the ranges should be chosen only within the ranges of the basis table for
each model. If they are outside these ranges error
messages\index{error message} will indicate the problems.

\subsection{Downloading and compiling}
The file {\tt composemodules.f90}\index{composemodules.f90}
contains all the 
necessary modules and the file {\tt compose.f90}\index{compose.f90} all required 
(sub-)routines and functions in order to read, interpolate and 
write the EoS data in ASCII format. If output of the EoS data 
in HDF5-format is needed, in addition the 
files {\tt hdf5compose.f90}\index{hdf5compose.f90} 
and {\tt hdf5writecompose.c}\index{hdf5writecompose.c} are required. 
All the files can be downloaded from the corresponding CompOSE 
web page\index{web page} and they 
have to be compiled with an 
appropriate {\sc Fortran90}\index{compiler!Fortran90} 
or {\sc C}\index{compiler!C} compiler.
The program was written using the GNU compilers\index{compiler!GNU} 
{\tt gfortran} and {\tt gcc}
and the use of these compilers is encouraged. A sample  
{\tt Makefile}\index{Makefile} is available 
on the CompOSE web page. It contains in line 26 a switch
to select the compilation without ({\tt HDF = 0}, default setting)
and with ({\tt HDF = 1}) the option for the HDF5 output. For using the HDF5 routines, 
the HDF5 library has to be installed. Compiling the
program files with the provided {\tt Makefile} generates an
executable with the name {\tt compose}.

\subsection{Direct use of {\tt compose}}

In this subsection, the details are described how customized
EoS tables from the EoS models on the
CompOSE web site can be obtained by supplying tables with the relevant
input values and running the program {\tt compose}\index{compose}. 

The operation proceeds in the following steps:

\begin{enumerate}
\item{\bf Selection of EoS}
\index{equation of state!selection}

In a first step the user selects the EoS that is appropriate for
her/his application from the list given on the CompOSE web pages.
Often, there are different types of tables available for a single
EoS model as described in subsection \ref{sec:tab_structure}.
The tabulated data can depend on one, two or
three of the independent parameters $T$, $n_{b}$ and $Y_{q}$. 
There are at least four different files needed for the successful 
application of the program {\tt compose}\index{compose}
as specified in subsection \ref{ssec:identtables}: three files
containing information on the parameter grid and at least one
file with the thermodynamic data.

\item{\bf Preparation of input files}
\index{file!input}

The downloaded parameter and EoS files are always identified with a
particular name and a file name extension that specifies the type
of stored quantities. The program {\tt compose}\index{compose}
uses generic names for the input files. Thus the downloaded files
have to be renamed such that the specific EoS name is changed to 
``{\tt eos}''
and the extension, e.g.\ ``{\tt .thermo}'', ``{\tt .compo}'', ``{\tt .micro}'',
``{\tt .t}'', ``{\tt .nb}'' or ``{\tt yq}'', is kept.
In addition to the downloaded files, the user has to supply two files
(in ASCII format)
that specify the parameter values and the quantities that will be
stored in the customized EoS output table. Generic files 
{\tt eos.parameters}\index{eos.parameters} and 
{\tt eos.quantities}\index{eos.quantities} can be downloaded from
the CompOSE web pages. They have the following structure:

\begin{itemize}
\item{\bf eos.parameters}

Rows one, three, five and seven of the file {\tt eos.parameters}\index{eos.parameters}
(in ASCII format) are
comment lines indicated by the first character {\tt \#}.
The second row of this file  
contains three integer numbers
$I_{T}$\index{$I_{T}$}, $I_{n_{b}}$\index{$I_{n_{b}}$} 
and $I_{Y_{q}}$\index{$I_{Y_{q}}$} that
specify the interpolation scheme for the temperature, baryon number
density and charge fraction of strongly interacting particles, 
respectively. Presently, there
is the choice between first order ($I_{x}=1$, $x=T,n_{b},Y_{q}$),
second order ($I_{x}=2$) 
and third order ($I_{x}=3$) interpolation\index{interpolation} available with
continuity of all quantities $Q(x)$, all quantities $Q(x)$ and their first
derivative $\partial Q/\partial x$ or all quantities $Q(x)$, their first and 
second derivatives $\partial Q/\partial x$ and $\partial^{2}
Q/\partial x^{2}$, respectively. See appendix~\ref{sec:interpol} for
details on the interpolation scheme.

The fourth row of the file {\tt eos.parameters} 
contains a single integer $I_{\beta}$\index{$I_{\beta}$} that determines whether the
EoS in the output file {\tt eos.table} will be generated for matter
in $\beta$ equilibrium ($I_{\beta} = 1$) or not ($I_{\beta} \neq 1$).
Of course, this option is only effective for EoS tables that include
electrons (and muons) and depend on the parameter $Y_{q}$, i.e.\
three-dimensional\index{table!three-dimensional} 
general purpose EoS tables and two-dimentional\index{table!two-dimensional}
zero-temperature EoS tables.

In the sixth row of the file, the tabulation scheme for the parameters
is defined by the integer $I_{\rm tab}$\index{$I_{\rm tab}$}. 
For $I_{\rm tab} = 0$ the the parameter values of $T$, $n_{b}$ and
$Y_{q}$
are listed explicitly in the file {\tt eos.parameters}\index{eos.parameters} as follows:
The eigth row contains the number of data points 
$N_{\rm data}$\index{$N_{\rm data}$} to be generated and
each line of the next $N_{\rm data}$ rows
contains the three parameter values in the form
\begin{equation}
 T \:\: n_{b} \:\: Y_{q}
\end{equation}
that define the grid points\index{grid point}.
The EoS data that will be stored in the final EoS table will appear
in the same sequence of the parameters as given in the file 
{\tt eos.parameters}\index{eos.parameters}.
For {$I_{\beta} =1$}, $Y_{q}$ can be set to any value.
For two- and one-dimensional EoS 
tables\index{table!two-dimensional}\index{table!one-dimensional}, 
the parameters that are not
used can be set to arbitrary values, e.q. for a zero-temperature EoS
the first entry in each row can have any finite value.
An example file {\tt eos.parameters} for the tabulation
  scheme with $I_{\rm tab}=0$ is given by
\begin{quote}
{\tt \# interpolation rules for T, n\_b and Y\_q} \\
{\tt 3 3 3} \\
{\tt \# calculation of beta equilibrium (1) or not (else)} \\
{\tt 0} \\
{\tt \# tabulation scheme of parameter values (see manual)} \\
{\tt 0} \\
{\tt \# parameter values depending on tabulation scheme} \\
{\tt 5} \\
{\tt 0.1 0.1 0.5} \\
{\tt 0.1 0.2 0.5} \\
{\tt 0.1 0.3 0.5} \\
{\tt 0.1 0.4 0.5} \\ 
{\tt 0.1 0.5 0.5} \\
\end{quote}
for generating a table using third order interpolations in $T$, $n_{b}$
and $Y_{q}$. Five data points are calculated for constant $T=0.1$~MeV and
$Y_{q} = 0.5$ at densities of $n_{b} = 0.1$, $0.2$, $0.3$, $0.4$, and $0.5$~fm$^{-3}$.

For $I_{\rm tab} \neq 0$ the parameter values are generated
in a similar way as suggested in subsection \ref{ssec:para}.
In this case, only three additional
rows follow the comment line below the row with the specification of $I_{\rm tab}$.
Each of these rows (in the order temperature, density, charge
fraction) contains the four quantities 
\begin{equation}
 p_{\rm min}\index{$p_{\rm min}$} \quad p_{\rm max}\index{$p_{\rm max}$} 
\quad N_{p}\index{$N_{p}$} \quad I_{p}\index{$I_{p}$}
\end{equation}
where $p$\index{$p$} stands for the parameters $T$, $n_{b}$ or $Y_{q}$, respectively.
The minimum and maximum values of the parameter $p$ in the EoS table
are denoted by $p_{\rm min}$ and $p_{\rm max}$, respectively. $N_{p}>0$
is the number of data points in the mesh of the parameter $p$ and
$I_{p}$ defines the discretization scheme. If $I_{p} = 0$, the
individual points are given by a linear interpolation
\begin{equation}
 p_{i}\index{$p_{i}$} = p_{\rm min} + (p_{\rm max} - p_{\rm min}) \frac{i-1}{N_{p}-1}
\end{equation}
for $i=1,\dots,N_{p}$ with $N_{p}>1$ and 
\begin{equation}
 p_{1} = p_{\rm min}
\end{equation}
for $N_{p}=1$. For $I_{p}\neq 0$, a logarithmic scaling is used with
\begin{equation}
 p_{i} =p_{\rm min} \left( \frac{p_{\rm max}}{p_{\rm min}}\right)^{\frac{i-1}{N_{p}-1}}
\end{equation}
for $i=1,\dots,N_{p}$ with $N_{p}>1$ and 
\begin{equation}
 p_{1} = p_{\rm min}
\end{equation}
for $N_{p}=1$. 
For this tabulation scheme an example file {\tt
    eos.parameters}
has the form
\begin{quote}
{\tt \# interpolation rules for T, n\_b and Y\_q} \\
{\tt 3 3 3} \\
{\tt \# calculation of beta equilibrium (1) or not (else)} \\
{\tt 0} \\
{\tt \# tabulation scheme of parameter values (see manual)} \\
{\tt 1} \\
{\tt \# parameter values depending on tabulation scheme} \\
{\tt 5.0 5.0 1 0} \\
{\tt 0.01 1.0 201 1} \\
{\tt 0.3 0.3 1 0}
\end{quote}
Here, $T=5$~MeV and $Y_{q}=0.3$ are kept constant and
the densities $n_{b}$ cover the range from $0.01$~fm$^{-3}$ to
$1.0$~fm$^{-3}$ in $200$ intervals with a logarithmic scaling.

In case the option $I_{\beta}=1$\index{$N_{\beta}$} is set in the third
row of the file {\tt eos.parameters}, the specification of the
parameter grid in $Y_{q}$ is irrelevant.

\item{\bf eos.quantities}
\index{eos.quantities}

This file contains 18 rows. They determine which
and how quantities will be stored in the final EoS table with the name
{\tt eos.table}\index{eos.table}.
Rows with odd number (one, three, \dots) are comment lines
beginning with the character {\tt \#}.
The second and fourth lines define the thermodynamic quantities that will be
stored in the output file {\tt eos.table}. The first entry 
$N_{\rm thermo}$\index{$N_{\rm thermo}$}  in row two specifies
the number of thermodynamic quantities that are selected from
table \ref{tab:ident_thermo}
followed by the number $N_{\rm add}$\index{$N_{\rm add}$} of quantities that are
selected from the stored quantities of the EoS file 
{\tt eos.thermo}\index{eos.thermo} in addition to
the seven standard quantities as described in subsection \ref{ssec:tabthermoquant}.
The fourth line of the file {\tt eos.quantities}\index{eos.quantities} lists the
$N_{\rm thermo}$ 
indices from the second column of table \ref{tab:ident_thermo}
that define the available thermodynamic quantities
and the $N_{\rm add}$ indices of the additionally stored quantities.
E.g., the first four lines
\begin{quote}
{\tt \# number of regular and additional thermodynamic quantities} \\
{\tt 3 1} \\
{\tt \# indices of regular and additional thermodynamic quantities} \\
{\tt 6 1 2 1}
\end{quote}
of the file {\tt eos.quantities}\index{eos.quantities}
denote that the quantities $\mathcal{F}/m_{n}-1$, $p$, $\mathcal{S}$
and $q_{1}$ will appear in the 
outpout file {\tt eos.table}\index{eos.table}
or {\tt eoscompose.h5}\index{eoscompose.h5}.
Regular thermodynamic quantities are those that are obtained by a direct
interpolation of the tabulated quantities $\mathcal{Q}_{1}$, \dots,
$\mathcal{Q}_{7}$ stored in the file {\tt eos.thermo} or those that are
obtained from them by applying the thermodynamic relations of
section \ref{sec:thermo_coeff}.
The indices of the regular thermodynamic quantities can have
values of $1$,
$2$, \dots, $19$ corresponding to the definition in table
\ref{tab:ident_thermo}.
Additional thermodynamic quantities are those that are denoted 
$q_{1}$, \dots, $q_{N_{\rm add}}$ in the file {\tt eos.thermo}. Their
number $N_{\rm add}$ and meaning depends on the specific EoS table.
Their values are found by a direct interpolation.
If both $N_{\rm thermo}$\index{$N_{\rm thermo}$} 
and $N_{\rm add}$\index{$N_{\rm add}$} are zero the fourth line of the
file {\tt eos.quantities}\index{eos.quantities} is empty.

The sixth row of the file {\tt eos.quantities}\index{eos.quantities} contains the numbers
$N_{p}$\index{$N_{p}$} and $N_{q}$\index{$N_{q}$} 
of pairs\index{pair} and quadruples\index{quadruple} that are selected from the
compositional quantities\index{quantity!compositional} 
that are stored in the file {\tt eos.compo}\index{eos.compo}
as described in subsection \ref{ssec:compo}. In the eigth line
the $N_{p}$ particle indices $I_{1}$, \dots, $I_{N_{p}}$ as defined in
tables \ref{tab:partindex} and \ref{tab:corrindex} 
are followed by the $N_{q}$ indices that 
identify a particular group of nuclei, see subsection
\ref{ssec:compo}.
E.g., the four lines
\begin{quote}
{\tt \# number of pairs and quadruples for composition data} \\
{\tt 3 1} \\
{\tt \# indices of pairs and quadruples for composition data} \\
{\tt 10 11 0 1}
\end{quote}
denote that the fractions $Y_{n}$, $Y_{p}$, $Y_{e}$ are stored in the
file {\tt eos.table}\index{eos.table} followed by 
$A_{1}^{\rm av}$\index{$A^{\rm av}$}, $Z_{1}^{\rm av}$\index{$Z^{\rm av}$}
and $Y_{1}^{\rm av}$\index{$Y^{\rm av}$} of the first set of nuclei.
If both $N_{p}$\index{$N_{p}$} and $N_{q}$\index{$N_{q}$} are zero the eighth line of the
file {\tt eos.quantities}\index{eos.quantities} is empty.

The tenth line of the file {\tt eos.quantities} contains the number
$N_{\rm mic}$\index{$N_{\rm mic}$} of
microscopic quantities\index{quantity!microscopic} in the file 
{\tt eos.micro}\index{eos.micro} to be
stored in the output file {\tt eos.table}\index{eos.table}. Then line 12
is a list of $N_{\rm mic}$ composite indices\index{index!composite} 
$K_{i}$\index{$K_{i}$} as defined in
subsection \ref{ssec:micro}.
E.g., the four lines
\begin{quote}
{\tt \# number of microscopic quantities} \\
{\tt 2} \\
{\tt \# indices of microscopic quantities} \\
{\tt 10050 11050}
\end{quote}
denote that the nonrelativistic single-particle potentials
$U_{n}$ and $U_{p}$ are stored in the file {\tt eos.table}.
If $N_{\rm mic}$  is zero the twelfth line of the
file {\tt eos.quantities} is empty.

In addition to the quantities considered 
above, error estimates\index{error!estimate} for the
interpolation\index{interpolation} 
of the thermodynamic
quantities $\mathcal{F}$\index{$\mathcal{F}$}, 
$\mathcal{E}$\index{$\mathcal{E}$}, 
$p/n_{b}$\index{$p$} and $\mathcal{S}$\index{$\mathcal{S}$} are available.
The free energy per baryon $\mathcal{F}$ can be obtained by direct
interpolation or by using the homogeneity\index{homogeneity} condition (\ref{eq:f_con})
with the interpolated values for the pressure and chemical potentials.
An estimate for the absolute error\index{error!absolute} 
in $\mathcal{F}$ is then given by
\begin{equation}
 \Delta \mathcal{F}\index{$\Delta \mathcal{F}$} = \mathcal{F} + \frac{p}{n_{b}}
 - \left( \mu_{b} + Y_{q} \mu_{q}\right)
\end{equation}
in case of an EoS without leptons and
\begin{equation}
 \Delta \mathcal{F} = \mathcal{F} + \frac{p}{n_{b}}
 - \left( \mu_{b} + Y_{l} \mu_{l}\right)
\end{equation}
in case of an EoS with leptons and charge neutrality.
The relative error\index{error!relative} is then given by 
$\Delta \mathcal{F}/\mathcal{F}-1$.
The consideration above also apply to the internal energy per baryon $\mathcal{E}$.
An estimate for the absolute error in $p/n_{b}$ is obtained by comparing
the directly interpolated pressure $p$ with that derived from the
first derivative of the free energy per baryon, i.e.\
\begin{equation}
 \Delta \frac{p}{n_{b}} = \frac{p}{n_{b}} 
 - n_{b} \left. \frac{\partial \mathcal{F}}{\partial n_{b}}
 \right|_{T,Y_{q}} \: .
\end{equation}
Similarly, an error estimate for the entropy per baryon is calculated
from
\begin{equation}
 \Delta \mathcal{S}\index{$\Delta \mathcal{S}$} = \mathcal{S}
 - \left. \frac{\partial \mathcal{F}}{\partial T}
 \right|_{n_{b},Y_{q}} \: .
\end{equation}
The sixteenth line of the file {\tt eos.quantities}\index{eos.quantities}
defines the number
$N_{\rm err}$\index{$N_{\rm err}$}
of error quantities to be stored in the output file.
Line 18 is a list of the $N_{\rm err}$ indices $J$\index{$J$}
as defined in table \ref{tab:ident_err}.
E.g., the four lines
\begin{quote}
{\tt \# number of error quantities} \\
{\tt 2} \\
{\tt \# indices of error quantities} \\
{\tt 1 2 }
\end{quote}
denote that the estimates
$\Delta \mathcal{F}$ and $\Delta \mathcal{F}/\mathcal{F}$ 
are stored in the file {\tt eos.table}\index{eos.table}.
If $N_{\rm err}$ is zero the eighteenth line of the
file {\tt eos.quantities} is empty.

The last row of the file {\tt eos.quantities}\index{eos.quantities}
defines the format\index{format!output} of
the output file  by a single integer $I$.  
There are two methods that are used for storing the EoS data. The
first ($I=1$) is the simple ASCII format\index{format!ASCII} 
using the file name {\tt eos.table}\index{eos.table}
that easily allows to read the data without the need for further
codes.
The second method ($I \neq 1$) 
uses the more advanced HDF5 format\index{format!HDF5} (see
\url{http://www.hdfgroup.org/HDF5/}) that is widely used in the
astrophysics community.  In this case, the output file carries the
name {\tt eoscompose.h5}\index{eoscompose.h5} and
each data set is designated with a
particular identifier\index{identifier} 
{\tt $ \$ $} of the stored 
quantity $Q$. Tables
\ref{tab:hdf5} and \ref{tab:hdf5micro} 
give a list of
all tabulated quantities and corresponding identifiers 
{\tt $ \$ $}\index{$ \$ $}.

Note that $\varepsilon=\mathcal{E}/m_{n}-1$\index{$\varepsilon$} 
in table~\ref{tab:ident_thermo}
is \emph{not} the specific internal
energy\index{energy!internal!specific}, 
since the total mass
density\index{density!mass!total} is not given by $m_{n} n_{b}$.
We do not store $\mathcal{E}/m_{n}$\index{$\mathcal{E}$} because it is in many regimes
largely dominated by the rest mass contribution, such that it is numerically
difficult to keep trace of small variations in the internal energy in that
case. The same argument applies to the other thermodynamic
quantities $\mathcal{F}$\index{$\mathcal{F}$}, 
$\mathcal{G}$\index{$\mathcal{G}$} and 
$\mathcal{H}$\index{$\mathcal{H}$}.
The present choice of storing $\varepsilon$ is also 
motivated by the fact that the total
mass density is not a conserved quantity\index{quantity!conserved} 
throughout the hydrodynamic
evolution, but only the baryon number density, $n_{b}$. 

The individual chemical
potentials\index{potential!chemical!individual} 
for all the particles (baryons\index{baryon},
mesons\index{meson}, nuclei\index{nucleus}, quarks\index{quark}, \dots)
can be calculated from the three ones given in the tables. Note that we
use the relativistic definition\index{definition!relativistic} 
of the chemical potentials.

Particle fractions\index{fraction!particle} $Y_{i}$\index{$Y_{i}$} 
are stored in the data sets with identifier\index{identifier}
$y\#$ where
$\#$\index{$\#$} stands for the particle index $I_{i}$ as defined in tables
\ref{tab:partindex} and \ref{tab:corrindex}. E.g., the data identifier
{\tt y20} and {\tt y400} 
contain the particle fractions of
$\Delta^{-}$ and $K^{-}$ particles, respectively.
Similarly, {\tt yav1} stands for the fraction of the first group of
nuclei.

Combining the examples above, the file {\tt eos.quantities} assumes the
form
\begin{quote}
{\tt \# number of regular and additional thermodynamic quantities} \\
{\tt 3 1} \\
{\tt \# indices of regular and additional thermodynamic quantities} \\
{\tt 6 1 2 1} \\
{\tt \# number of pairs and quadruples for composition data} \\
{\tt 3 1} \\
{\tt \# indices of pairs and quadruples for composition data} \\
{\tt 10 11 0 1} \\
{\tt \# number of microscopic quantities} \\
{\tt 2} \\
{\tt \# indices of microscopic quantities} \\
{\tt 10050 11050} \\
{\tt \# number of error quantities} \\
{\tt 2} \\
{\tt \# indices of error quantities} \\
{\tt 1 2 }
\end{quote}
and, consequently,
each line of the output file {\tt eos.table} 
will contain the following entries:
{
\begin{eqnarray}
 \nonumber 
T \quad n_{b} \quad Y_{q} & & 
\mathcal{F}/m_{n}-1 \quad
 p \quad \mathcal{S} \quad q_{1} \quad Y_{n} \quad Y_{p} \quad Y_{e}
 \quad 
\\
 \nonumber  & & 
  A^{\rm av}_{1} \quad Z^{\rm av}_{1}  \quad 
 Y^{\rm av}_{1}  \quad U_{n} \quad U_{p} \quad \Delta \mathcal{F} \quad \Delta
 \mathcal{F}/\mathcal{F} 
\end{eqnarray}
}
i.e., in total $17$ quantities.

\end{itemize}

\item{\bf Running {\tt compose}}

After starting the {\tt compose} program, the relevant input data
files are read: first the files with the parameter grids
{\tt eos.t}\index{eos.t}, {\tt eos.nb}\index{eos.nb} 
and {\tt eos.yq}\index{eos.yq}, second the EoS data files
{\tt eos.thermo}\index{eos.thermo}, {\tt eos.compo}\index{eos.compo} 
(if available) and {\tt eos.micro}\index{eos.micro}
(if available). The read data are analyzed, checked for
consistency and a report of the results in written to the file
{\tt eos.report}\index{eos.report} that is also used to extract data for the
EoS data sheet\index{data sheet} provided on the web page. 
Next, the user supplied input files
{\tt eos.parameters}\index{eos.parameters} 
and {\tt eos.quantities}\index{eos.quantities} are read and the output
file {\tt eos.table}\index{eos.table} is generated.

Each row of the output file {\tt eos.table} contains 
the three parameter values of the user-chosen grid as defined in the file
{\tt eos.parameters} and then the selected quantities as specified and
in the order of the file {\tt eos.quantities}. Note that for each
quadruple index\index{index!quadruple} $I_{i}$ in the fourth row of the file 
{\tt  eos.quantities} three quantities ($A^{\rm av}_{I_{i}}$, 
$Z^{\rm av}_{I_{i}}$ and $Y_{I_{i}}$) are given. 
See the example given above.

\end{enumerate}

\subsection{Using subroutines and modules}

Instead of using the code {\tt compose.f90} as given, the user can
employ the subroutines directly in her/his program. 
The file {\tt compose.f90}\index{compose.f90} 
contains one main program,
24 subroutines\index{subroutine} and one
function. 
The structure of the program with the dependencies of the subroutines
and functions is
depicted in tables \ref{tab:code} and \ref{tab:code2}.
The file {\tt composemodules.f90}\index{composemodules.f90}
contains two modules\index{module}. In the module {\tt eos\_tables}
the dimensions {\tt dim\_t=101}\index{dim\_t}, 
{\tt dim\_n=401}\index{dim\_n}, and {\tt dim\_y=100}\index{dim\_y} for the
maximum values $N_{T}^{\rm max}$\index{$N_{T}^{\rm max}$}, 
$N_{n_{b}}^{\rm max}$\index{$N_{n_{b}}^{\rm max}$}, and
$N_{Y_{q}}^{\rm max}$\index{$N_{Y_{q}}^{\rm max}$}, respectively, are
defined, see subsection \ref{ssec:para}. They can be adjusted to
larger values if required. Note that
{\tt dim\_a}\index{dim\_a} has to be set to the maximum value of 
{\tt dim\_t}, {\tt dim\_n}, and {\tt dim\_y}. 

The user needs to call the following three subroutines 
\begin{quote}
 {\tt init\_eos\_table(iwr)} \\
 {\tt define\_eos\_table(iwr)} \\
 {\tt get\_eos(t,n,y,ipl\_t,ipl\_n,ipl\_y)}
\end{quote}
from the file {\tt compose.f90} in order to
generate EoS data. The first two subroutines
depend on a single integer parameter {\tt iwr}.
For {\tt iwr = 1} a progress report during the execution of the
subroutine will be written to the terminal. Otherwise, this action will be
suppressed. The subroutine {\tt get\_eos} has three {\tt double
  precision} parameters ({\tt t}, {\tt n}, {\tt y}) 
and three {\tt integer} parameters ({\tt ipl\_t}, {\tt ipl\_n}, {\tt ipl\_y})
as arguments, see below.
Be sure to include all (sub)routines, functions and the two
modules {\tt eos\_tables} and {\tt compose\_internal} in your code.

The subroutine {\tt init\_eos\_table} has to be called only once 
in order to initialize all relevant tables and quantities. The files
{\tt eos.t}, {\tt eos.nb}, {\tt eos.yq} and
{\tt eos.thermo} have to exist in order that {\tt init\_eos\_table}
is executed properly. The files {\tt eos.compo} and {\tt eos.micro}
are optional.

The subroutine\index{subroutine} {\tt define\_eos\_table} defines the quantities that
are interpolated and that will be stored in the EoS
table. It reads the file {\tt eos.quantities}. The subroutine needs to
be called only once. Instead of calling 
{\tt define\_eos\_table} in the user's program, it is possible that
the relevant parameters are defined directly in the user's code. 
Thus it is possible to change the selection of quantities that
are interpolated during the execution of the user's program.
In the following, the variable names to be specified by the user are
given according to the structure of the input file 
{\tt eos.quantities}\index{eos.quantities}:
\begin{quote}
 {\tt \# number of regular and additional thermodynamic quantities} \\
 {\tt n\_qty} \quad {\tt n\_add} \\
 {\tt \# indices of regular and additional thermodynamic quantities} \\
 {\tt idx\_qty(1)} \quad \dots \quad
 {\tt idx\_qty(n\_qty)} \quad 
 {\tt idx\_add(1)} \quad \dots \quad
 {\tt idx\_add(n\_add)} \\
 {\tt \# number of pairs and quadruples for composition data} \\
 {\tt n\_p} \quad {\tt n\_q} \\
 {\tt \# indices of pairs and quadruples for composition data} \\
 {\tt idx\_p(1)} \quad \dots \quad
 {\tt idx\_p(n\_p)} \quad 
 {\tt idx\_q(1)} \quad \dots \quad
 {\tt idx\_q(n\_q)} \\
 {\tt \# number of microscopic quantities} \\
 {\tt n\_m} \\
 {\tt \# indices of microscopic quantities} \\
 {\tt idx\_m(1)} \quad \dots \quad
 {\tt idx\_m(n\_m)} \\
 {\tt \# number of error quantities} \\
 {\tt n\_err} \\
 {\tt \# indices of error quantities} \\
 {\tt idx\_err(1)} \quad \dots \quad
 {\tt idx\_err(n\_err)} \\
 {\tt \# format of output file} \\
 {\tt iout}
\end{quote}
Note that all variables are {\tt integers}. The appearing vectors
are defined in the module\index{module} {\tt compose\_internal}:
\begin{quote}
 {\tt idx\_qty(dim\_qtyt)} \quad \mbox{with} \quad {\tt dim\_qtyt = 19} \\
 {\tt idx\_add(dim\_qty)}  \quad \mbox{with} \quad {\tt dim\_qty = 15}\\
 {\tt idx\_p(dim\_qtyp)} \quad \mbox{with} \quad {\tt dim\_qtyp = 10}\\
 {\tt idx\_q(dim\_qtyq)} \quad \mbox{with} \quad {\tt dim\_qtyq = 2} \\
 {\tt idx\_m(dim\_qtym)} \quad \mbox{with} \quad {\tt dim\_qtym = 16} \\
 {\tt idx\_err(dim\_qtye)} \quad \mbox{with} \quad {\tt dim\_qtye = 8} \: .
\end{quote}
where the dimensions are defined in the module {\tt eos\_tables}.

Finally, the subroutine\index{subroutine} 
{\tt get\_eos} can be called as often as needed
with the appropriate arguments. The {\tt double precision} variables
{\tt t}, {\tt n} and {\tt y} define the temperature $T$ [MeV], baryon
number density $n_{b}$ [fm${}^{-3}$] and the charge fraction
of strongly interacting particles
$Y_{q}$ [dimensionless], respectively, where the EoS is evaluated.
The {\tt integer} indices {\tt ipl\_t}, {\tt ipl\_n} and
{\tt ipl\_y} define the interpolation rule in $T$, $n_{b}$ and
$Y_{q}$, respectively. For {\tt ipl\_t = 1} all selected quantities
are interpolated linearly such that the interpolated values agree
with the tabulated values at the grid points of the EoS table.
For {\tt ipl\_t = 2} the interpolated values of a quantity and
its first derivative with respect to temperature agree with those
at the grid points. For {\tt ipl\_t = 3} a continuity of also the
second derivatives at the grid points are demanded. Similarly, the
interpolation rules for $n_{b}$ and $Y_{q}$ are determined. If the
index is outside the range $[1,3]$ it is set to $3$.
See appendix~\ref{sec:interpol}  for details of the interpolation scheme.

The results of the interpolation are stored in five {\tt double
  precision} vectors/arrays with dimensions defined in the module\index{module}
{\tt eos\_tables}:
\begin{quote}
  {\tt eos\_thermo(dim\_qtyt)}  \quad \mbox{with} \quad {\tt dim\_qtyt = 19}\\
  {\tt eos\_thermo\_add(dim\_qty)} \quad \mbox{with} \quad {\tt dim\_qty = 15}\\
  {\tt eos\_compo\_p(dim\_qtyp)} \quad \mbox{with} \quad {\tt dim\_qtyp = 10}\\
  {\tt eos\_compo\_q(dim\_qtyq,3)} \quad \mbox{with} \quad {\tt dim\_qtyq = 2} \\  
  {\tt eos\_micro(dim\_qtym)} \quad \mbox{with} \quad {\tt dim\_qtym = 16} \\
  {\tt eos\_err(dim\_qtye)} \quad \mbox{with} \quad {\tt dim\_qtye =
    8} \: .
\end{quote}
The vector index in {\tt eos\_thermo} corresponds to the index $J$
given in table~\ref{tab:ident_thermo}. The vector index in {\tt
  eos\_thermo\_add} is just the index $i=1,\dots,N_{\rm add}$ of the 
additional quantities stored in each row of the file {\tt eos.thermo}.
The (first) index of the vectors/arrays {\tt eos\_compo\_p}, 
{\tt eos\_compo\_q} and {\tt eos\_compo\_m} corresponds to the index
of the vectors {\tt idx\_p}, {\tt idx\_q} and {\tt idx\_m} that are
defined in the file {\tt eos.quantities} or by the user before
the subroutine {\tt get\_eos} is called. The second index
of the array {\tt eos\_compo\_q} is correlated with the index $J$ as
given in table \ref{tab:ident_compo}.
The vector index in {\tt eos\_err} corresponds to the index $J$
given in table~\ref{tab:ident_err}.

\subsection{HDF5 table}
\index{table!HDF5}
The organization of the data in case of the HFD5 table is different
as compared to the ASCII table. All data are stored in a single
data file that is denominated {\tt eoscompose.h5}\index{eoscompose.h5}.
The names of the data sets are given by the
identifiers\index{identifier!HDF5} in tables
\ref{tab:hdf5} and \ref{tab:hdf5micro}. Note, too, 
that the values of the different quantities at all grid points 
are stored together in one data set. The data sets can 
in addition contain a group of quantities, e.g., 
the $N_{\mathrm{thermo}}$ thermodynamic quantities, 
see tables~\ref{tab:hdf5} and \ref{tab:hdf5micro} for details. 

\subsection{Information for data sheet}
\index{data sheet}

During the execution of the program {\tt compose} a file
{\tt eos.report}\index{eos.report} is generated that contains information on the
EoS derived from the tabulated data. The information is used for the
creation of the data sheet\index{data sheet} that accompanies each EoS
on the CompOSE web pages\index{web page}. If possible, a second file
{\tt eos.beta}\index{eos.beta} is produced with the EoS of $\beta$-equilibrated
matter for the lowest available temperature.
Each row consists of four quantities, the density $n_{b}$, the charge
fraction $Y_{q}$, the (free) energy density $f$ (including rest mass
contributions) and the pressure $p$.
These data are used to derive characteristic neutron star parameters
for the data sheet.

\subsection{Error messages}
\index{error message}

During the execution of the subroutines, errors can occur due to
several reasons, e.g.\ parameters out of range etc. In this case,
the execution is stopped and a corresponding error message is generated.

\begin{table}[htb]
\begin{center}
\caption{\label{tab:ident_thermo}%
Thermodynamic quantities\index{quantity!thermodynamic} 
which are stored in the data tables
with their units. }
{\small 
\begin{tabular}{cccc}
\toprule
index & quantity/  & unit & description \\ 
$J$& expression & & \\
\toprule
1 & $p$\index{$p$} & MeV~fm${}^{-3}$ & total pressure \\ 
\midrule
2 & $\mathcal{S}$\index{$\mathcal{S}$} 
& dimensionless & total entropy per baryon \\ 
\midrule
 3 & $\mu_{b}-m_{n}$\index{$\mu_{b}$} & MeV & baryon chemical potential \\ 
                    &     & & with respect to neutron mass \\
\midrule
4 & $\mu_{q}$\index{$\mu_{q}$} & MeV & 
 {\small charge chemical potential} \\ 
\midrule
5 & $\mu_{l}$\index{$\mu_{l}$} & MeV & lepton chemical potential \\ 
\midrule
6 & $\mathcal{F}/m_{n} -1$\index{$\mathcal{F}$} & dimensionless 
       & free energy per baryon   \\  
        & & & with respect to neutron mass  \\ 
\midrule
 7 & $\varepsilon =$\index{$\varepsilon$} & dimensionless 
       & internal energy per baryon   \\  
       &  $\mathcal{E}/m_{n} -1$\index{$\mathcal{E}$} 
& &  with respect to neutron mass  \\ 
\midrule
8 & $\mathcal{H}/m_{n}-1$\index{$\mathcal{F}$} & dimensionless 
       & enthalpy per baryon \\
       & & & with respect to neutron mass \\ 
\midrule
9 & $\mathcal{G}/m_{n}-1$\index{$\mathcal{G}$} & dimensionless 
       & free enthalpy per baryon \\
        & & & with respect to neutron mass \\ 
\midrule
10 & $\left. \frac{\partial p}{\partial n_{b}} \right|_{\mathcal{E}}$ & MeV 
       & {\small partial derivative of pressure with}\\ 
        & & & {\small respect to baryon number density}\\ 
\midrule
11 & $\left. \frac{\partial p}{\partial
    \mathcal{E}}\right|_{n_{b}}$ & fm$^{-3}$
       & {\small partial derivative of pressure}\\ 
        & & & {\small with respect to internal} \\
        & & & {\small energy per baryon} \\
\midrule
12 & $c_{s}^{2}$\index{$c_{s}$} & dimensionless 
    & speed of sound squared \\ 
\midrule
13 & $c_{V}$\index{$c_{V}$} & dimensionless & specific heat capacity \\
            &               & & at constant volume \\ 
\midrule
14 & $c_{p}$\index{$c_{p}$} & dimensionless & specific heat capacity \\
            &               & & at constant pressure \\ 
\midrule
15 & $\Gamma = c_{p}/c_{V}$\index{$\Gamma$} & dimensionless & adiabatic index \\
\midrule
16 & $\alpha_{p}$\index{$\alpha_{p}$} & MeV${}^{-1}$ & expansion coefficient \\
                    &              & & at constant pressure \\
\midrule
17 & $\beta_{V}$\index{$\beta_{V}$} & fm${}^{-3}$ & tension coefficient \\
                   &            & & at constant volume \\ 
\midrule
18 & $\kappa_{T}$\index{$\kappa_{T}$} 
& fm${}^{3}$/MeV & isothermal compressibility\\ 
\midrule
19 & $\kappa_{S}$\index{$\kappa_{S}$} 
& fm${}^{3}$/MeV & adiabatic compressibility \\ 
\bottomrule
\end{tabular}
}
\end{center}
\end{table}

\begin{table}[th]
\begin{center}
\caption{\label{tab:ident_compo}%
Quantities containing information on the composition\index{quantity!compositional}
which are stored in the data tables
with their units.
The particle index $I_{i}$\index{$I_{i}$} has been
defined in tables \ref{tab:partindex} and \ref{tab:corrindex}. 
The symbols \% denotes the group of nuclei.} 
{\small
\begin{tabular}{cccc}
\toprule
index $J$ & quantity/  & unit & description \\ 
 & expression  & & \\
\toprule
 & $I_{\rm phase}$\index{$I_{\rm phase}$} 
& dimensionless & phase index \\
\midrule
 & $Y_{I_{i}}$\index{$Y_{i}$} & dimensionless & fraction of particle $I_{i}$ \\
\midrule
$1$ & $Y^{\rm av}_{\%}$\index{$X^{\rm av}$} 
& dimensionless & combined fraction  \\
   & & & of group \% of nuclei \\
\midrule
$2$ & $A^{\rm av}_{\%}$\index{$A^{\rm av}$} 
& dimensionless & average mass number  \\
   & & & of group \% of nuclei
\\
\midrule
$3$ & $Z^{\rm av}_{\%}$\index{$Z^{\rm av}$} 
& dimensionless & average charge number \\
  & & & of group \% of nuclei \\
\bottomrule
\end{tabular}
}
\end{center}
\end{table}

\begin{table}[htb]
\begin{center}
\caption{\label{tab:ident_micro}%
Microscopic quantities\index{quantity!microscopic}
which are stored in the data tables
with their unit.}
{\small
\begin{tabular}{cccc}
\toprule
index $J$ & quantity/ & unit & description \\ 
 & expression & & \\
\toprule
40 & $m^{L}_{I_{i}}/m_{I_{i}}$\index{$m^{L}_{I_{i}}$} 
& dimensionless & effective Landau mass \\ 
 & & & with respect to \\
 & & & the particle mass \\
\midrule
41 & $m^{D}_{I_{i}}/m_{I_{i}}$\index{$m^{D}_{I_{i}}$}
& dimensionless & effective Dirac mass \\ 
 & & & with respect to \\
 & & & the particle mass \\
\midrule
50 & $U_{I_{i}}$\index{$U_{I_{i}}$} & MeV & nonrelativistic  \\ 
 & & & single-particle potential \\
\midrule
51 & $V_{I_{i}}$\index{$V_{I_{i}}$} & MeV & relativistic  \\ 
  & & & vector self-energy \\
\midrule
52 & $S_{I_{i}}$\index{$S_{I_{i}}$} & MeV & relativistic  \\ 
  & & & scalar self-energy \\
\midrule
60 & $\Delta_{I_{i}}$\index{$\Delta_{I_{i}}$} & MeV & gap \\
\bottomrule
\end{tabular}
}
\end{center}
\end{table}

\begin{table}[htb]
\begin{center}
\caption{\label{tab:ident_err}%
Error quantities which are stored in the data tables
with their unit.}
{\small
\begin{tabular}{cccc}
\toprule
index $J$ & quantity/  & unit & description \\ 
 & expression & & \\
\toprule
1 & $\Delta \mathcal{F}$\index{$\Delta \mathcal{F}$} 
& MeV & absolute error estimate \\ 
   & & of free energy per baryon \\
\midrule
 2 & $\Delta \mathcal{F}/\mathcal{F}$ & [dimensionless]
 & relative error estimate \\ 
  & & of free energy per baryon \\
\midrule
 3 & $\Delta \mathcal{E}$\index{$\Delta \mathcal{E}$} 
& MeV & absolute error estimate \\ 
  & & of internal energy per baryon \\
\midrule
 4 & $\Delta \mathcal{E}/\mathcal{E}$ & [dimensionless]
 & relative error estimate \\ 
  & & of internal energy per baryon \\
\midrule
 5 & $\Delta \left(\frac{p}{n_{b}}\right)$ & MeV 
 & absolute error estimate \\ 
  & & of pressure-to-density ratio \\
\midrule
 6 & $\Delta \left(\frac{p}{n_{b}}\right)/
 \left(\frac{p}{n_{b}}\right)$ & [dimensionless] 
 & relative error estimate \\ 
  & & of pressure-to-density ratio \\
\midrule
 7 & $\Delta \mathcal{S} $\index{$\Delta \mathcal{S}$} & [dimensionless]
 & absolute error estimate \\ 
  & & of entropy per baryon \\
\midrule
 8 & $\Delta \mathcal{S}/\mathcal{S}$ & [dimensionless] 
 & relative error estimate \\ 
  & & of entropy per baryon \\
\bottomrule
\end{tabular}
}
\end{center}
\end{table}

\begin{table}[htdp]
\begin{center}
\caption{\label{tab:code}%
Modules\index{module}, (sub)routines\index{subroutine} 
and functions in the file {\tt compose.f90}\index{compose.f90} with their
dependencies (continued on next page).}
{\small 
\begin{tabular}{llll}
\toprule
  & uses {\tt MODULE} & uses {\tt SUBROUTINE} & uses {\tt FUNCTION} \\
\toprule
 {\tt PROGRAM} & & & \\
\toprule
 {\tt compose} & & {\tt init\_eos\_table} & \\
 & & {\tt define\_eos\_table} & \\
 & & {\tt get\_eos\_table} & \\
\toprule
 {\tt SUBROUTINE} & & & \\
\toprule
 {\tt init\_eos\_table} & & {\tt
   read\_eos\_tables\_tny} \\
 & & {\tt read\_eos\_table\_thermo} & \\
 & & {\tt read\_eos\_table\_compo} & \\
 & & {\tt read\_eos\_table\_micro} & \\
 & & {\tt get\_diff\_rules} & \\
 & & {\tt init\_ipl\_rule} & \\
 & & {\tt get\_eos\_report} & \\
\midrule
 {\tt define\_eos\_table} & {\tt compose\_internal} & {\tt
   write\_errors} & \\
\midrule
 {\tt get\_eos\_table} & {\tt compose\_internal} & {\tt get\_eos} & \\
 & & {\tt write\_errors} & \\
\midrule
 {\tt get\_eos} & {\tt compose\_internal} & {\tt get\_eos\_beta} & \\
 & & {\tt get\_eos\_sub} & \\
 & & {\tt write\_errors} & \\
\midrule
 {\tt get\_eos\_beta} & {\tt compose\_internal} & {\tt get\_eos\_sub} & \\
\midrule
 {\tt get\_eos\_sub} & {\tt compose\_internal} & {\tt
   get\_eos\_grid\_para} & \\
 & & {\tt eos\_interpol} & \\
\midrule
 {\tt get\_eos\_grid\_para} & {\tt compose\_internal} & & \\
\midrule
 {\tt eos\_interpol} & {\tt compose\_internal} & {\tt
   get\_idx\_arg1} & {\tt get\_ipl\_rule} \\ 
 & & {\tt get\_idx\_arg2} & \\
 & & {\tt get\_diff\_rules2} & \\
 & & {\tt get\_interpol\_yq} & \\
 & & {\tt get\_interpol\_nb} & \\
 & & {\tt get\_derivatives} & \\
 & & {\tt get\_coefficients} & \\
 & & {\tt get\_interpol\_xy} & \\
\midrule
 {\tt get\_idx\_arg1} & {\tt compose\_internal} & & \\
\midrule
 {\tt get\_idx\_arg2} & {\tt compose\_internal} & & \\
\bottomrule
\end{tabular} 
}
\end{center}
\end{table}

\begin{table}[htdp]
\begin{center}
\caption{\label{tab:code2}%
Modules\index{module}, (sub)routines\index{subroutine} 
and functions in the file {\tt compose.f90}\index{compose.f90} with their
dependencies (continued from previous page).}
{\small 
\begin{tabular}{llll}
\toprule
  & uses {\tt MODULE} & uses {\tt SUBROUTINE} & uses {\tt FUNCTION} \\
\toprule
 {\tt SUBROUTINE} & & & \\
\toprule
 {\tt get\_diff\_rules} & {\tt compose\_internal} & & \\
\midrule
 {\tt get\_diff\_rules2} & {\tt compose\_internal} & & {\tt get\_ipl\_rule} \\
\midrule
 {\tt get\_interpol\_yq} & {\tt compose\_internal} & & \\
\midrule
 {\tt get\_interpol\_nb} & {\tt compose\_internal} & & \\
\midrule
 {\tt get\_interpol\_xy} & {\tt compose\_internal} & & \\
\midrule
 {\tt get\_derivatives} & {\tt compose\_internal} & & \\
\midrule
 {\tt get\_coefficients} & {\tt compose\_internal} & & \\
\midrule
 {\tt init\_ipl\_rule} & {\tt compose\_internal} & & \\
\midrule
 {\tt get\_eos\_report} & {\tt compose\_internal} & {\tt get\_eos\_nmp} & \\
 & & {\tt get\_eos} & \\
 & & {\tt write\_errors} & \\
\midrule
 {\tt get\_eos\_nmp} & {\tt compose\_internal} & {\tt get\_eos\_sub} & \\
\midrule
 {\tt write\_errors} & {\tt compose\_internal} & & \\
\midrule
 {\tt read\_eos\_tables\_tny} & {\tt compose\_internal} & {\tt
   write\_errors} & \\
\midrule
 {\tt read\_eos\_table\_thermo} & {\tt compose\_internal} & {\tt
   write\_errors} & \\
\midrule
 {\tt read\_eos\_table\_compo} & {\tt compose\_internal} & {\tt
   write\_errors} & \\
\midrule
 {\tt read\_eos\_table\_micro} & {\tt compose\_internal} & {\tt
   write\_errors} & \\
\toprule
 {\tt FUNCTION} & & & \\
\toprule
 {\tt get\_ipl\_rule} & & & \\
\bottomrule
\end{tabular} 
}
\end{center}
\end{table}

\newpage

\begin{table}[th]
\begin{center}
  \caption{\label{tab:hdf5}%
    Thermodynamical quantities 
    which are stored in the HDF5 data file together with their
    units and the name\index{identifier!HDF5} of the corresponding
    data set. If there are two lines in the 
    second column, the first one always
    corresponds to $I_{\mathrm{tab}} = 0$ and the second one to
    $I_{\mathrm{tab}} \neq 0$.
}
\begin{tabular}{cccc}
\toprule
name of  & quantity/  & unit & description \\ 
data set &  expression
 & & \\
\toprule
t & $T$\index{$T$} 
     & MeV & temperature \\ 
\midrule
pointst  &  $\begin{array}{c} N_{\rm data}\index{$N_{\rm data}$} \\ 
                               N_{T}\index{$N_{T}$} \end{array}$
       & dimensionless & number of points in $T$\\ 
\midrule
nb  & $n_{b}$\index{$n_{b}$}  
       & fm$^{-3}$ & baryon number density\\
\midrule 
pointsnb  &  $\begin{array}{c} N_{\rm data}  \\ 
                               N_{n_{b}}\index{$N_{n_{b}}$} \end{array}$
       & dimensionless & number of points in $n_{b}$\\ 
\midrule
yq    & $Y_{q}$\index{$Y_{q}$} & dimensionless & charge
fraction of strongly \\ 
 & & & interacting particles \\
\midrule
pointsyq  &  $\begin{array}{l} N_{\rm data}  \\ 
                               N_{Y_{q}}\index{$N_{Y_{q}}$}  \end{array}$
       & dimensionless & number of points in $Y_{q}$\\ 
\midrule
thermo    & 
  $Q(N_{\rm data},1,1,N_{\rm thermo})$
  & varying & array of
thermodynamical \\ &$Q(N_{n_{b}}, N_{T},N_{Y_{q}},N_{\rm thermo})$\index{$Q$}
&& quantities, see table~\ref{tab:ident_thermo} \\
\midrule
pointsthermo  &  $N_{\rm thermo}$\index{$N_{\rm thermo}$} 
       & dimensionless & number of thermodynamical \\ 
 &&& quantities, see table~\ref{tab:ident_thermo}.\\ 
\midrule
index\_thermo  &  $J(N_{\rm thermo})$\index{$J$}
       & dimensionless & index identifying the \\ &&& thermodynamical
       quantities, \\ &&& see table~\ref{tab:ident_thermo}.\\ 
\midrule
thermo\_add  & $q(N_{\rm data},1,1,N_{\rm add})$
& varying & array of additional \\ 
 &   $q(N_{n_{b}},N_{T},N_{Y_{q}},N_{\rm add})$\index{$q$} 
 & & quantities $q_{\%}$ \\ &&& in file {\tt
eos.thermo}. \\
\midrule
pointsadd  &  $N_{\rm add}$\index{$N_{\rm add}$}
       & dimensionless & number of additional \\ &&& thermodynamical
       quantities \\ &&& in file {\tt eos.thermo}.\\ 
\midrule
index\_thermo\_add  & $I_{\rm add}(N_{\rm add})$\index{$I_{\rm add}$} & dimensionless &
index identifying the \\ &&& additional quantities \\ & & & $q_{\%}$ from  file {\tt
eos.thermo},\\ &&& 
see eq.\ (\ref{eq:eos.thermo}) \\
\bottomrule
\end{tabular} 
\end{center}
\end{table}

\begin{table}[th]
\begin{center}
  \caption{\label{tab:hdf5micro}%
    Compositional and microscopic quantities 
    which are stored in the HDF5 data file together with their
    units and the name\index{identifier!HDF5} of the corresponding
    data set.
    If there are two lines in the 
    second column, the first one always
    corresponds to $I_{\mathrm{tab}} = 0$ and the second one to
    $I_{\mathrm{tab}} \neq 0$.
  }
\begin{tabular}{cccc}
\toprule
name of  & quantity/  & unit & description \\ 
data set &  expression
 & & \\
\toprule
yi    & $\begin{array}{l}
  Y_{I}(N_{\rm data},1,1,N_{\rm p})\index{$Y_{I}$} \\ 
  Y_{I}(N_{n_{b}}, N_{T},N_{Y_{q}},N_{\rm p})
  \end{array} $& dimensionless & array of
particle fractions \\ 
\midrule
pointspairs  &  $N_{\rm p}$\index{$N_{\rm p}$} 
       & dimensionless & number of pairs \\ & & & for compositional data \\  
\midrule
index\_yi  &  $I_{i}(N_{\rm p})$\index{$I_{i}$} 
       & dimensionless & index identifying the \\ &&& particle $i$,
       see table~\ref{tab:partindex}.\\ 
\midrule
yav    & $
  Y^{\rm av}(N_{\rm data},1,1,N_{\rm q})$\index{$Y^{\rm av}$}
&dimensionless & array of combined fractions \\
&  $Y^{\rm av}(N_{n_{b}}, N_{T},N_{Y_{q}},N_{\rm q})$
&& of groups of nuclei  \\ 
\midrule
aav    & $
  A^{\rm av}(N_{\rm data},1,1,N_{\rm q})$\index{$A^{\rm av}$}
&dimensionless & average mass number \\
& $ A^{\rm av}(N_{n_{b}}, N_{T},N_{Y_{q}},N_{\rm q}) $
&& of groups of nuclei  \\ 
\midrule
zav    & $
  Z^{\rm av}(N_{\rm data},1,1,N_{\rm q}) $\index{$Z^{\rm av}$}
&dimensionless & average charge number\\
&$  Z^{\rm av}(N_{n_{b}}, N_{T},N_{Y_{q}},N_{\rm q})$
&& of groups of nuclei  \\ 
\midrule
pointsav  &  $N_{\rm q}$\index{$N_{\rm q}$} 
       & dimensionless & number of quadruples  \\ &&&for compositional data \\  
\midrule
index\_av  &  $I_{\rm av}(N_{\rm q})$\index{$I_{\rm av}$} 
       & dimensionless & index identifying the group $\%$\\ & & &
       of nuclei, see table~\ref{tab:partindex}.\\ 
\midrule
micro    & $\begin{array}{l}
  q_{\rm mic}(N_{\rm data},1,1,N_{\rm mic})\index{$q_{\rm mic}$}  \\ 
  q_{\rm mic}(N_{n_{b}}, N_{T},N_{Y_{q}},N_{\rm mic}) 
 \end{array} $&varying & array of microscopic quantities\\ 
\midrule
pointsmicro  &  $N_{\rm mic}$\index{$N_{\rm mic}$} 
       & dimensionless & number of microscopic quantities\\  
\midrule
index\_micro  &  $K(N_{\rm mic})$\index{$K$}
       & dimensionless & index identifying the microscopic\\ & & &
       quantities, see subsection~\ref{ssec:micro}.\\ 
\midrule
error    & $\begin{array}{l}
  q_{\rm err}(N_{\rm data},1,1,N_{\rm err})\index{$q_{\rm err}$}  \\ 
  q_{\rm err}(N_{n_{b}}, N_{T},N_{Y_{q}},N_{\rm err }) \end{array} 
 $&varying & array of error quantities \\ 
\midrule
pointserr  &  $N_{\rm err}$\index{$N_{\rm err}$}
       & dimensionless & number of microscopical quantities\\  
\midrule
index\_err  &  $J_{\rm err}(N_{\rm err})$\index{$J_{\rm err}$} 
       & dimensionless & index identifying the error\\ & & &
       quantities, see table~\ref{tab:ident_err}.\\ 
\bottomrule
\end{tabular} 
\end{center}
\end{table}


\part{Appendix}
\appendix

\chapter{Technical details}
\label{app:techdetails}

The basic EoS tables only give a selected set of
thermodynamic quantities\index{quantity!thermodynamic} 
at the grid points\index{grid point} 
that are identified with the index triple\index{index!triple}
($i_{T}\index{$i_{T}$},i_{n_{b}}\index{$i_{n_{b}}$},i_{Y_{q}}$\index{$i_{Y_{q}}$}).
>From the stored quantities\index{quantity!stored} 
$Q_{i}$\index{$Q_{i}$}, $i=1,\dots,6$, see section
\ref{ssec:tabthermoquant}, further relevant thermodynamic quantities
can be derived by a smooth interpolation for all possible values
of the parameters\index{parameter} $T$, $n_{b}$ and $Y_{q}$ within the tabulated
ranges. For these quantities, thermodynamic 
consistency\index{consistency!thermodynamic} should be
respected as far as possible. When the thermodynamic
quantities are interpolated separately, however, this condition
is usually not exactly fulfilled. On the other hand, a separate 
interpolation often leads to smoother dependencies of the
thermodynamic quantitities on the parameters avoiding unphysical
oscillations. In addition, different ways of determining a single
quantity gives the opportunity to estimate the error in the
interpolation.
Thus, in the present code {\tt compose.f90}\index{compose.f90} the strategy of a
direct interpolation\index{interpolation} of individual quantities is followed.

\section{Interpolation}
\label{sec:interpol}

The interpolation\index{interpolation} scheme for a  
quantity in the thermodynamic, compositional and microscopic
EoS tables
generally proceeds in several steps.
In the following, the procedure will be explained for a
three-dimensional general
purpose EoS\index{equation of state!general purpose} table\index{table!three-dimensional}. 
The interpolation scheme follows
the method proposed in Ref.\ \cite{Swe96}
using polynomials of sufficiently high order.


In the sequel, the generic symbol $Q$\index{$Q$} will be used for any tabulated
quantity\index{quantity!tabulated}. 
Its values are given at the grid points\index{grid point} 
that are specified
by a triple of indices\index{index!triple} 
$(i_{T},i_{n_{b}},i_{Y_{q}})=(i,j,k)$, see subsection
\ref{ssec:para}, corresponding to temperature, baryon number density and
charge fraction of strongly interacting particles. 
Thus all values $Q(T(i),n_{b}(j),Y_{q}(k))$\index{$Q$} are known at the grid points. 
In order to calculate the
quantity $Q$ at given values of $T$, $n_{b}$, and $Y_{q}$, first the values
of the indices
$i = i_{T}\index{$i$}\index{$i_{T}$}$, 
$j = i_{n_{b}}\index{$j$}\index{$i_{n_{b}}$}$ and 
$k  = i_{Y_{q}}\index{$k$}\index{$i_{Y_{q}}$}$ are determined
such that
\begin{eqnarray}
 T(i)        \leq & T\index{$T$} & <  T (i+1) \: , \\
 n_{b}(j) \leq & n_{b}\index{$n_{b}$} & <  n_{b}(j+1) \: , \\
 Y_{q}(k) \leq & Y_{q}\index{$Y_{q}$} & <  Y_{q}(k+1) \: .
\end{eqnarray}
Then the interpolation parameters\index{parameter!interpolation}
\begin{eqnarray}
 \xi\index{$\xi$} & = & \frac{T-T(i)}{T(i+1)-T(i)} \\
 \eta\index{$\eta$} & = &
 \frac{n_{b}-n_{b}(j)}{n_{b}(j+1)-n_{b}(j)} \\
 \zeta\index{$\zeta$} & = &
 \frac{Y_{q}-Y_{q}(k)}{Y_{q}(k+1)-Y_{q}(k)} 
\end{eqnarray}
are introduced with
\begin{equation}
 0 \leq \xi < 1 \qquad
 0 \leq \eta < 1 \qquad
 0 \leq \zeta < 1 \: .
\end{equation}
For the interpolation of a quantity $Q$ at given $(T,n_{b},Y_{q})$
we need the tabulated values at least at the eight corners of the
cube\index{cube} with grid points 
$(i,j,k)$,
$(i+1,j,k)$,
$(i,j+1,k)$,
$(i,j,k+1)$,
$(i+1,j+1,k)$,
$(i+1,j,k+1)$,
$(i,j+1,k+1)$,
$(i+1,j+1,k+1)$.

The interpolation proceeds in two steps: first, an interpolation
in the variable $Y_{q}$ such that the three-dimensional grid\index{grid!three-dimensional}
is mapped to a two-dimensional grid\index{grid!two-dimensional} 
with four corners of each square;
second, a two-dimensional
interpolation\index{interpolation!two-dimensional} 
in the variables $T$ and $n_{b}$.

For the interpolation in $Y_{q}$ in the three-dimensional
cube\index{cube} as defined above, four separate one-dimensional 
interpolations\index{interpolation!one-dimensional} along the
lines that connect the grid points 
$(i,j,k)$ and $(i,j,k+1)$,
$(i+1,j,k)$ and $(i+1,j,k+1)$,
$(i,j+1,k)$ and $(i,j+1,k+1)$,
$(i+1,j+1,k)$ and $(i+1,j+1,k+1)$
have to be performed.
After this first step, the interpolation proceeds in the parameters $T$ and
$n_{b}$ by a two-dimensional\index{interpolation!two-dimensional} 
scheme as discussed in subsection 
\ref{subsec:2dim}.

The interpolation in the variables $T$, $n_{b}$ and $Y_{q}$ can
be of different order\index{interpolation!order} in general, defined by the value $I$ of the
variables {\tt ipl\_t}, {\tt ipl\_n} and {\tt ipl\_y}, see section
\ref{sec:data}. In the following, the case of highest order ($I=3$) is
considered first.

\subsection{Interpolation in one dimension}
\index{interpolation!one-dimensional}

The order $I=3$ of the interpolation requires
that the function values, its first and second derivatives have to be
continuous at the two corner points of each line, i.e.\ six values have to be
reproduced by the interpolation polynomial. Hence, a polynomial of
at least fifth degree has to be used. 
The six independent coefficients $q_{n}$ of a single
polynomial
\begin{equation}
\label{eq:poly}
 Q(T(i),n_{b}(j),Y_{q}) = \sum_{n=0}^{5} q_{n} \zeta^{n}
\end{equation}
can determined from the function values and derivatives\index{derivative} at the corner
points directly. One finds
\begin{equation}
 q_{0}\index{$q_{i}$} =  Q_{ijk}^{(0)} \qquad
 q_{1} = Q_{ijk}^{(1)} \qquad
 q_{2} = \frac{1}{2} Q_{ijk}^{(2)}
\end{equation}
with
\begin{equation}
 Q^{(n)}_{ijk}\index{$Q^{(n)}_{ijk}$}  = 
 \left[Y_{q}(k+1)-Y_{q}(k)\right]^{n}\left. \frac{\partial^{n}Q}{\partial
   Y_{q}^{n}}\right|_{T(i),n_{b}(j),Y_{q}(k)}   
\end{equation}
for $n=0,1,2$, and
\begin{equation}
 Q_{ijk}^{(0)} = \left. \frac{\partial^{0}Q}{\partial
     Y_{q}^{0}}\right|_{T(i),n_{b}(j),Y_{q}(k)}
 = Q(T(i),n_{b}(j),Y_{q}(k))
\end{equation}
in particular. The remaining three coefficients are given by
\begin{eqnarray}
 q_{3} & = & 10 A -4 B + \frac{1}{2} C \\ 
 q_{4} & = & -15 A + 7 B - C \\
 q_{5} & = & 6 A - 3 B + \frac{1}{2}C
\end{eqnarray}
with
\begin{eqnarray}
 A\index{$A$} & = &  Q_{ijk+1}^{(0)}- Q_{ijk}^{(0)} - Q_{ijk}^{(1)} - \frac{1}{2}
 Q_{ijk}^{(2)} \\
 B\index{$B$} & = & Q_{ijk+1}^{(1)} - Q_{ijk}^{(1)} - Q_{ijk}^{(2)} \\
 C\index{$C$} & = & Q_{ijk+1}^{(2)} - Q_{ijk}^{(2)} \: .
\end{eqnarray}

Alternatively, the approach of Ref.\
\cite{Swe96} can be followed.
The value of the quantity $Q$ at given $\zeta$ is found 
with the help of the quintic basis functions\index{basis function!quintic}
\begin{eqnarray}
 \psi_{0}^{(0)}(z) & = & 1 - 10 z^{3} + 15 z^{4} - 6 z^{5}
 \\
 \psi_{1}^{(0)}(z) & = & z - 6 z^{3} + 8 z^{4} - 3 z^{5}
 \\
 \psi_{2}^{(0)}(z) & = & \frac{1}{2} \left( z^{2} - 3 z^{3} + 3 z^{4} -
   z^{5} \right) 
\end{eqnarray}
that have the properties
\begin{equation}
 \psi_{m}^{(n)}(0)  =  \delta_{nm} \qquad
 \psi_{m}^{(n)}(1)  =  0
\end{equation}
for 
\begin{equation}
 \psi_{m}^{(n)}(z)\index{$\psi_{m}^{(n)}$} = \frac{d^{n}\psi_{m}^{(0)}}{dz^{n}}
\end{equation}
with $n,m \in \{0,1,2\}$.
Then one has
\begin{eqnarray}
\label{eq:quintic_z}
 Q(T(i),n_{b}(j),Y_{q}) & = &
 \sum_{n=0}^{2} \left[ Q^{(n)}_{ijk} \psi_{n}^{(0)}(\zeta) +
 (-1)^{n} Q^{(n)}_{ijk+1} \psi_{n}^{(0)}(1-\zeta)\right]
\end{eqnarray}
with six interpolation coefficients\index{interpolation!coefficient}
that are given by
directly by the function values and derivatives\index{derivative}
at the corner points defined by the indices. 
Derivatives are easily found as
\begin{eqnarray}
 \lefteqn{\frac{\partial^{s}}{\partial Y_{q}^{s}} Q(T(i),n_{b}(j),Y_{q})}
 \\ \nonumber & = &
 \frac{1}{\left[Y_{q}(k+1)-Y_{q}(k)\right]^{s}}
 \sum_{n=0}^{2} \left[ Q^{(n)}_{ijk} \psi_{n}^{(s)}(\zeta) +
 (-1)^{n+s} Q^{(n)}_{ijk+1} \psi_{n}^{(s)}(1-\zeta)\right] \: .
\end{eqnarray}

The numerical determination of the derivatives 
depends on the order of interpolation. For $I=3$, centered five-point
finite difference formulas\index{formula!finite difference} 
are used to calculate the first and second
derivative\index{derivative} of a function. Close to the boundaries of the EoS,
centered difference formulas cannot be applied. In this case,
off-center formulas are used. In the case $I=2$, three-point 
finite difference formulas are employed.
A continuity of the
second derivatives at the corner points is not required. Only the
function and the first derivative are demanded to be continuous,
corresponding to four independent quatities. Hence, a polynomial of
third degree is suffiecient and $q_{4} = q_{5} = 0$ in Eq.\ (\ref{eq:poly}).
Then, the coefficients of the second derivatives have to be defined as
\begin{eqnarray}
 Q_{ijk}^{(2)} & = & 6 \left[ Q_{ijk+1}^{(0)} - Q_{ijk}^{(0)} \right]
 - 2 Q_{ijk+1}^{(1)} - 4 Q_{ijk}^{(1)}
 \\ 
 Q_{ijk+1}^{(2)} & = & - 6 \left[ Q_{ijk+1}^{(0)} - Q_{ijk}^{(0)} \right]
 + 4 Q_{ijk+1}^{(1)} + 2 Q_{ijk}^{(1)} \: .
\end{eqnarray}
For $I=1$, the function has to be continuous at the grid points
but there is no condition on the derivatives. In this case,
one sets 
\begin{equation}
 Q_{ijk}^{(1)} = Q_{ijk+1}^{(1)} = Q_{ijk+1}^{(0)}-Q_{ijk}^{(0)} \qquad
 Q_{ijk}^{(2)} = Q_{ijk+1}^{(2)} = 0 \
\end{equation}
for given function values $Q_{ijk}^{(0)}$ and $Q_{ijk+1}^{(0)}$.
Then the polynomial (\ref{eq:poly}) reduces to a linear function
with $q_{2} = q_{3} = q_{4} = q_{5} = 0$.

\subsection{Interpolation in two dimensions}
\label{subsec:2dim}
\index{interpolation!two-dimensional}

The interpolation in two dimensions with conditions on the
continuity of the function and its derivatives is more complicated
than in the one-dimensional case. For an interpolation
order\index{interpolation!order} $I=3$ in
both variables $\xi$ and $\eta$, there are 
four function values,
eight first derivatives and the twelve second derivatives, hence
24 values in total,
that can be used to determine the coefficients of a polynomial
in $\xi$ and $\eta$.  From this consideration, it would seem to be
sufficient to use a polynomial 
\begin{equation}
 Q(T,n_{b},Y_{q}) = \sum_{m=0}^{4} \sum_{n=0}^{4} q_{mn} \xi^{m} \eta^{n}
\end{equation}
that includes powers up to four in both
variables, resulting in $5 \times 5 = 25$ coefficients $q_{nm}$
in total, i.e.\ one more than
required. However, such a form does not guarantee that the function and the
derivatives are continuous not only at the grid points but also along
the boundaries of the interpolation square. In fact, a continuity of
the function $Q$, the first derivatives\index{derivative} $\partial Q/\partial T$ and
$\partial Q/\partial n_{b}$, the second derivatives 
$\partial^{2}Q/\partial T^{2}$,
$\partial^{2}Q/\partial T \partial n_{b}$ and
$\partial^{2}Q/\partial n_{b}^{2}$,
the third derivatives
$\partial^{3}Q/\partial T^{2} \partial n_{b}$ and
$\partial^{3}Q/\partial T \partial n_{b}^{2}$ and
the fourth derivative
$\partial^{4}Q/\partial T^{2} \partial n_{b}^{2}$ at the corners
has to be demanded, determining $4\times 9 = 36 = 6\times 6$
coefficients of a polynomial 
\begin{equation}
\label{eq:bipoly}
 Q(T,n_{b},Y_{q}) = \sum_{m=0}^{5} \sum_{n=0}^{5} q_{mn} \xi^{m} \eta^{n}
\end{equation}
with degree six in each variable.

Instead of determining the coefficients directly as in the case 
of a one dimensional interpolation, it is more advantageous
to use the biquintic
interpolation\index{interpolation!biquintic} 
scheme as in Ref.\ \cite{Swe96} with
\begin{eqnarray}
\label{eq:biquintic_xy}
 \lefteqn{Q(T,n_{b},Y_{q})}
 \\ \nonumber & = &
 \sum_{m=0}^{2} \sum_{n=0}^{2} 
 \left[ Q^{(mn)}_{ij} \psi_{m}^{(0)}(\xi)\psi_{n}^{(0)}(\eta)
 + (-1)^{m} Q^{(mn)}_{i+1j} \psi_{m}^{(0)}(1-\xi)\psi_{n}^{(0)}(\eta)
 \right. \\ \nonumber & & \left.
 + (-1)^{n} Q^{(mn)}_{ij+1} \psi_{m}^{(0)}(\xi)\psi_{n}^{(0)}(1-\eta)
 + (-1)^{m+n} Q^{(mn)}_{i+1j+1} \psi_{m}^{(0)}(1-\xi)\psi_{n}^{(0)}(1-\eta) \right]
\end{eqnarray}
and $4 \times 9 = 36$ coefficients
\begin{eqnarray}
Q^{(mn)}_{ij}\index{$Q_{ij}^{(mn)}$}
& = & \left[T(i+1)-T(i)\right]^{m}\left[n_{b}(j+1)-n_{b}(j)\right]^{n}
 \left. \frac{\partial^{m+n} Q}{\partial T^{m} \partial n_{b}^{n}} 
 \right|_{T(i),n_{b}(j),Y_{q}}  \\
Q^{(mn)}_{i+1j} & = & 
\left[T(i+1)-T(i)\right]^{m}\left[n_{b}(j+1)-n_{b}(j)\right]^{n}
 \left. \frac{\partial^{m+n} Q}{\partial T^{m} \partial n_{b}^{n}} 
 \right|_{T(i+1),n_{b}(j),Y_{q}}  \\
Q^{(mn)}_{ij+1} & = & 
\left[T(i+1)-T(i)\right]^{m}\left[n_{b}(j+1)-n_{b}(j)\right]^{n}
 \left. \frac{\partial^{m+n} Q}{\partial T^{m} \partial n_{b}^{n}} 
 \right|_{T(i),n_{b}(j+1),Y_{q}}  \\
Q^{(mn)}_{i+1j+1} 
 & = & \left[T(i+1)-T(i)\right]^{m}\left[n_{b}(j+1)-n_{b}(j)\right]^{n}
 \left. \frac{\partial^{m+n} Q}{\partial T^{m} \partial n_{b}^{n}} 
 \right|_{T(i+1),n_{b}(j+1),Y_{q}} 
\end{eqnarray}
where $m,n \in \{0,1,2\}$. Of course, it is possible to express the
coefficients $q_{mn}$\index{$q_{mn}$} of the polynomial (\ref{eq:bipoly}) through the
coefficients $Q_{ij}^{(mn)}$\index{$Q_{ij}^{(mn)}$}, 
as done in the code {\tt compose.f90}.
The derivatives are obtained in the same way as in
the one-dimensional case for the different interpolations orders $I$.
Derivatives\index{derivative} of the function $Q$ with respect to $Y$ and $n_{b}$ can be
found with the help of the relation
\begin{eqnarray}
 \lefteqn{\frac{\partial^{s+t}}{\partial T^{s}\partial n_{b}^{t}}Q(T,n_{b},Y_{q})}
 \\ \nonumber & = &
 \frac{1}{\left[T(i+1)-T(i)\right]^{s}}
 \frac{1}{\left[n_{b}(j+1)-n_{b}(j) \right]^{t}}
 \\ \nonumber & & \times
 \sum_{m=0}^{2} \sum_{n=0}^{2} 
 \left[ Q^{(mn)}_{ij} \psi_{m}^{(s)}(\xi)\psi_{n}^{(t)}(\eta)
 + (-1)^{m+s} Q^{(mn)}_{i+1j} \psi_{m}^{(s)}(1-\xi)\psi_{n}^{(t)}(\eta)
 \right. \\ \nonumber & & \left.
 + (-1)^{n+t} Q^{(mn)}_{ij+1} \psi_{m}^{(s)}(\xi)\psi_{n}^{(t)}(1-\eta)
 + (-1)^{m+n+s+t} Q^{(mn)}_{i+1j+1}
 \psi_{m}^{(s)}(1-\xi)\psi_{n}^{(t)}(1-\eta) \right] \: .
\end{eqnarray}

Above considerations apply to the interpolation in all three
parameters for a general purpose EoS 
table\index{table!general purpose}. 
For neutron matter\index{matter!neutron}, i.e.\
$Y_{q}=0$ the interpolation in the charge fraction of strongly
interacting particles is
trivial since $\zeta=0$ and the tabulated values of the quantities
can be used directly in the second interpolation step in $T$ and
$n_{b}$. A similar procedure applies to the case of an EoS for
symmetric nuclear matter\index{matter!symmetric} of $\beta$-equilibrium.
In the case of a zero-temperature table, the second
interpolation step reduces from two to one dimension.

\section{Matter in $\beta$ equilibrium}
\label{sec:betaequi}
\index{matter!$\beta$ equilibrium}

In the case of matter in $\beta$
equilibrium\index{equilibrium!$\beta$}, 
there is an additional
condition on the chemical potentials\index{chemical potential} that reduces the number of
independent parameters\index{parameter!independent} 
by one. Requiring that the weak
interaction reactions\index{reaction!weak interaction}
\begin{equation}
  p + e^{-} \leftrightarrow n + \nu_{e} \qquad
  p + \bar{\nu}_{e} \leftrightarrow n + e^{+}
\end{equation}
and, if muons are considered in the model,
\begin{equation}
  p + \mu^{-} \leftrightarrow n + \nu_{\mu} \qquad
   p + \bar{\nu}_{\mu} \leftrightarrow n + \mu^{+}
\end{equation}
are in equilibrium, the relations
\begin{equation}
 \mu_{p} + \mu_{e} = \mu_{n} + \mu_{\nu_{e}}
\end{equation}
and
\begin{equation}
 \mu_{p} + \mu_{\mu} = \mu_{n} + \mu_{\nu_{\mu}} \: ,
\end{equation}
respectively, should hold. 
The chemical potentials of the
neutrinos\index{potential!chemical!neutrino} are considered to be zero, i.e.\
$\mu_{\nu_{e}}=\mu_{\nu_{\mu}}=0$. 
Because $\mu_{p} = \mu_{n}+\mu_{q}$,
$\mu_{e} = \mu_{le}-\mu_{q}$ and $\mu_{\mu} = \mu_{l\mu}-\mu_{q}$,
the constraint can be formulated as
\begin{equation}
 \mu_{l}\index{$\mu_{l}$} = 0
\end{equation}
with the effective lepton chemical potential\index{potential!chemical!lepton}
$\mu_{l} = \mu_{le} = \mu_{l\mu}$. For given temperature $T$
and baryon number density $n_{b}$ there is a unique charge
fraction of strongly interacting particles
$Y_{q}$ that is found by determining the zero of the function
\begin{equation}
 f(Y_{q}) = \mu_{l}(Y_{q}) \: .
\end{equation}

\chapter{Organization of the CompOSE team}
\label{app:team}
\index{team!organization}
CompOSE is developed in close collaboration with the communities 
the service is meant for.
These are mainly physicists who develop equations of state or
simulate astrophysical phenomena numerically on the computer.
Direct interaction with the users is an essential part in the concept of the
project.




There is a \emph{core team}\index{team!core} 
that is engaged in the preparation of
the web site\index{web site}, the manual\index{manual} 
and the tools, numerical codes and tables. Members of the core team are
\setkomafont{labelinglabel}{\bf}
\begin{labeling}{FirstName LongFamilyName}
\item[Thomas Kl\"ahn]U Wroc{\l}aw (Poland),
\item[Micaela Oertel]LUTH Meudon (France),
\item[Stefan Typel]GSI Darmstadt (Germany). 
\end{labeling}

The core team is surrounded by a group of closely collaborating
physicists that contribute to various aspects of
the CompOSE project, e.g. in preparing EoS tables for particular
models or testing the CompOSE web site and tools. 
Presently, the \emph{support team}\index{team!support} consists of



\setkomafont{labelinglabel}{\bf}
\begin{labeling}{FirstName LongFamilyName}
\item[David Blaschke]JINR Dubna (Russia), U Wroc{\l}aw (Poland),
\item[Tobias Fischer]GSI Darmstadt (Germany), U Wroc{\l}aw (Poland),
\item[Matthias Hempel]U Basel (Switzerland),
\item[Daniel Zab{\l}ocki] U Wroc{\l}aw (Poland).
\end{labeling}

\chapter{Acknowledgments}


CompOSE would not be possible without the financial and organisatorial
support from a large number of institutions and individual contributors.
We gratefully acknowledge support by CompStar, a Research Networking Program
of the European Science Foundation (ESF) and the birthplace of
the CompOSE project, by a grant from the Polish Ministry for
Science and Higher Education (MNiSW) supporting the "CompStar"-activity,
by the Instytut Fizyki Teoretycznej of the Uniwersytet Wroc\l{}awski,
the National Science Centre Poland
(Narodowe Centrum Nauki, NCN) within the ``Maestro''-programme under
contract No. DEC-2011/02/A/ST2/00306, by the
``hadronphysics3'' network within the seventh framework program of the
European Union,
by the GSI Helmholtzzentrum f\"{u}r Schwerionenforschung GmbH,
by the Helmholtz International
Center for FAIR within the framework of the LOEWE program launched
by the state of Hesse via the Technical University Darmstadt,
by the Helmholtz Association (HGF) through the Nuclear Astrophysics
Virtual Institute (VH-VI-417),
by the ExtreMe Matter Institute EMMI in the framework
of the Helmholtz Alliance `Cosmic Matter in the Laboratory',
by the DFG cluster of excellence ``Origin and Structure of
the Universe'' and
by the SN2NS project ANR-10-BLAN-0503.


\newpage
\addcontentsline{toc}{chapter}{Index}
\printindex


\end{document}